\documentclass{jfm}

\usepackage{graphicx}
\usepackage{float}
\usepackage{newtxtext}
\usepackage{amssymb}
\usepackage{color}
\usepackage{newtxmath}
\usepackage{natbib}
\usepackage{hyperref}
\usepackage{epsfig,subfigure}
\usepackage{amsmath}
\usepackage{amssymb}
\usepackage[utf8]{inputenc}
\graphicspath{{figures/}}
\usepackage{cleveref}
\crefname{section}{§}{§§}
\Crefname{section}{§}{§§}
\usepackage[mathscr]{eucal}
\usepackage{float}
\hypersetup{
    colorlinks = true,
    urlcolor   = blue,
    citecolor  = black,
}

\newcommand{\RomanNumeralCaps}[1]
\linenumbers

\title{The bistability of curved compression ramp flows}

\author{Yan-Chao Hu\aff{1,2}
	\corresp{\email{huyanchao@cardc.cn}},
	Wen-Feng Zhou\aff{1,2},
	Ming-Zhi Tang\aff{1,2},
	Gang Wang\aff{1,2},
	Ming Fang\aff{1,2},
	Yan-Guang Yang\aff{1,3}\corresp{\email{yangyanguang@cardc.cn}},
	\and Zhi-Gong Tang\aff{3}
}

\affiliation{\aff{1}Laboratory of Aerodynamics in Multiple Flow Regimes, China Aerodynamics Research and Development Centre (CARDC), Mianyang 621000, China
	\aff{2}Hypervelocity Aerodynamics Institute, CARDC, Mianyang 621000, China
	\aff{3}CARDC, Mianyang 621000, China}
\begin{document}
	\maketitle
	
\begin{abstract}
\par This paper investigates the bistability of curved compression ramp (CCR) flows. It reports that both separated and attached states can be stably established even for the same boundary conditions, revealing that the ultimate stable states of CCR flows also depend on the initial conditions and evolutionary history. Firstly, we design a thought experiment involving two establishment routes of CCR flows, constructing two distinct-different stable states, respectively. Subsequently, three-dimensional direct numerical simulations are meticulously performed to instantiate the thought experiment, verifying the existence of the bistable states. Finally, we compare the pressure, wall friction, and heat flux distributions under two stable states. As a canonical type of Shock wave-Boundary layer interactions, local CCR flows often appear on aircraft, hence the bistability will certainly bring noteworthy changes to the global aerothermodynamic characteristics, which supersonic/hypersonic flight has to deal with.
\end{abstract}


\section{Introduction}
\label{sec:Introduction}
\par In shock wave-boundary layer interaction (SBLI) flows, separation and attachment of boundary layers are the two most typical states \citep{babinsky2011shock}. The ultimate stable states of a specific SBLI flow are widely considered to be unique for given inflow parameters and wall geometry, i.e., the boundary conditions. Although empirical, this understanding is still correct in most cases. However, in principle, effects of initial condition and evolutionary history of SBLIs can not be ignored, especially considering the reciprocal causation of shock wave (SW) patterns and boundary layer behaviors. An extreme question stemming from these process-dependent effects is that ``can both stable separated and attached states exist for the same boundary conditions in \textit{pure} SBLI flows?'' The statement ``\textit{pure}'' means that the flow is dominated by self-organized SBLIs rather than other multistable interactions, such as SW reflections \citep{ben2001hysteresis,tao2014viscous}. If the answer is ``yes'', the corresponding bistability will certainly bring significant differences to the aerothermodynamic characteristics, which is inevitable for super/hypersonic flight. Herein, a class of \textit{pure} SBLI flows, the canonical compression ramp (CR) flows, are chosen to investigate the process-dependent effects, and their bistability is reported.
\par Large separated CR flows, classified as type VI of SW interactions by \citet{edney1968anomalous}, are composed of large separation bubbles and $\lambda$-shock patterns. As a typical wall geometry, direct CRs (DCRs, the inclined ramp directly connecting the flat plate with one apex) are investigated widely, including the interaction processes with boundary layers' distortions near separation \citep{chapman1958investigation,stewartson1969self,neiland1969theory}, the aerodynamic characteristics \citep{gumand1959on,hung1973interference,simeonides1995experimental,tang2021aerothermodynamic}, the vortex structures inside both the separation bubbles \citep{gai2019hypersonic,cao2021unsteady} and the evolving boundary layers \citep{fu2021shock,hu2017beta}, and the unsteadiness of the SBLIs \citep{ganapathisubramani2009low,helm2021characterization,cao2021unsteady}. On the other hand, due to the potential to weaken separation, curved CR (CCR, inclined ramp connecting the flat plate with curved walls) flows have gradually attracted the attentions of researchers \citep{tong2017direct,wang2019amplification,hu2020bistable}. However, as far as we know, there are few reports about the multistability of \textit{pure} SBLIs in the CR flows.
\par The rest of this paper is organized as follows. Sec. \ref{sec:Thought experiment} describes a thought experiment to construct two distinct-different stable CCR flows for the same boundary conditions. Sec. \ref{sec:Numerical expriment} presents the three-dimensional (3D) direct numerical simulations (DNSs) to instantiate the thought experiment and verify the CCR flows' bistability. Conclusions follow in Sec. \ref{sec:Conclusions}.
\section{Thought experiment}
\label{sec:Thought experiment}
\begin{figure}
	\centering
	{\includegraphics[width = 0.95\columnwidth]{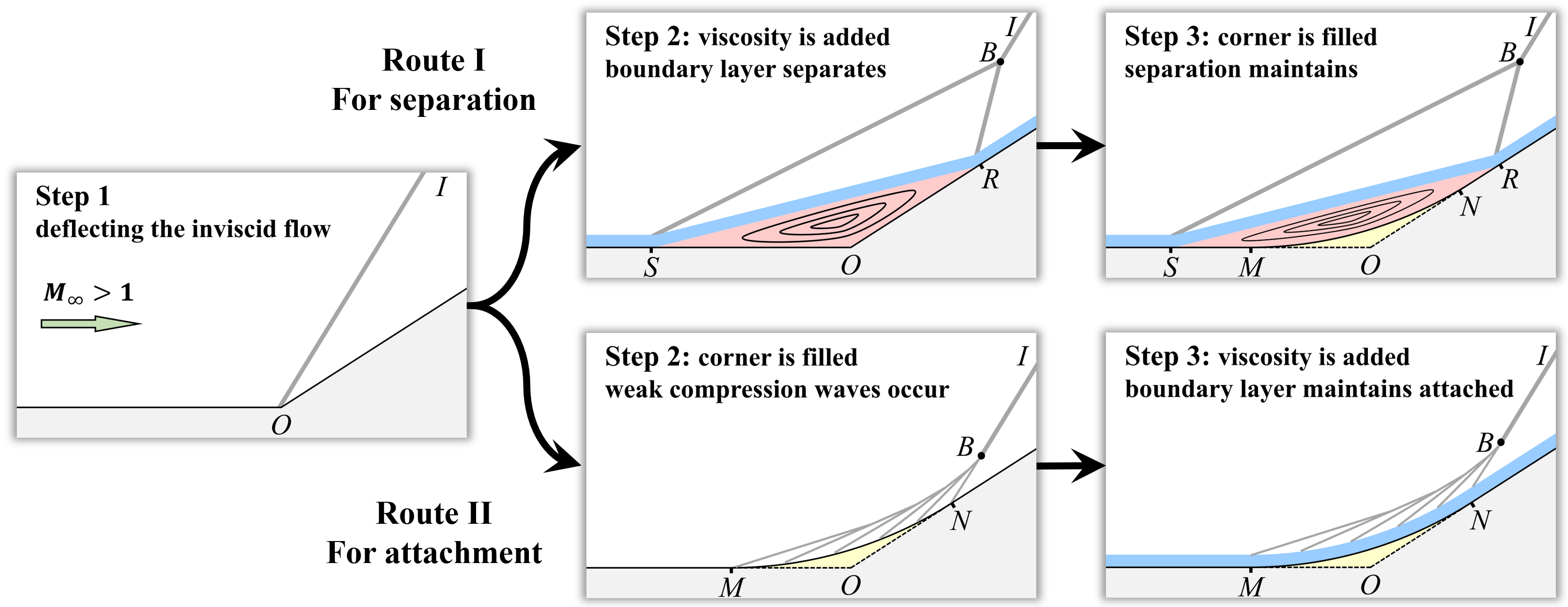}}
	\caption{Two possible distinct-different (separation and attachment) stable CCR flows constructed via Route I and II.}
\label{fig:two_routes}
\end{figure}

In this section, we design a thought experiment to construct two distinct-different stable CCR flows, separation and attachment, via Route I and II, respectively, as shown in Fig. \ref{fig:two_routes}.
\par First, a possible separated CCR flow is constructed via Route I with three steps.
\begin{itemize}
	\item {\bf Step 1: deflect the inviscid flow.} Set the initial flow field as an inviscid supersonic flow being deflected by a DCR geometry with ramp angle $\alpha$. An oblique SW $OI$ is formed and emits from the ramp apex $O$;
	\item {\bf Step 2: add the viscosity.} At some instant, the viscosity is suddenly added to the fluid. An attached boundary layer is subsequently formed and interacts with SW $OI$. For a large enough $\alpha$, the attached boundary layer can not resist the strong adverse pressure gradient induced by SW $OI$ and then separates from the wall. The inverse flow gradually shapes a closed separation bubble $SRO$, deflecting both the attached boundary layer and the outer inviscid flow, and inducing two new SWs, $SB$ and $RB$. As time goes on, the separation bubble stops growing and stabilizes, i.e., both the separation point $S$ and the reattachment point $R$ become almost fixed or only oscillate slightly;
	\item {\bf Step 3: fill the corner into an arc.} The key to filling is to be slow, gentle and macroscopically continuous, ensuring that the flow is always in stable states during the filling process. Thus, we fill the corner $O$ with an `imaginary machine' that can produce solid wall material atom by atom. In order to minimize the disturbance, the time span between two fillings is longer than the relaxation time of the flow stabilization. In the end, the corner $O$ is filled into an arc $\overline{MN}$ (the yellow region) inside space $SRO$, and a stably separated CCR flow could be obtained, as shown in the end of Route I in Fig. \ref{fig:two_routes}.
\end{itemize}

\par Second, a possible attached CCR flow is constructed via Route II using three steps.
\begin{itemize}
	\item {\bf Step 1: deflect the inviscid flow.} This step is the same as Step 1 in Route I;
	\item {\bf Step 2: fill the corner into an arc.} This step is similar to Step 3 in Route I, i.e., filling the corner into an arc (the yellow region). The only difference is that the corner's filling comes before the fluid viscosity's adding in Route II. Due to the curved wall $\overline{MN}$, SW $OI$ emitting from point $O$ is weakened to a series of compression waves spreading on $\overline{MN}$;
	\item {\bf Step 3: add the viscosity.} Similar to Step 2 in Route I, the viscosity is suddenly added to the fluid after the inviscid flow stably passing over the CCR in this step. Undergoing the weaker adverse pressure gradients of compression waves on $\overline{MN}$, the gradually formed boundary layer is very likely to stabilize at attached state, as shown in the end of Route II.
\end{itemize}
\par Compare Route I and II. Operationally, both stable separated and attached CCR flows could be obtained with three steps by only exchanging the order of their last two steps. However, in essence, the different presentations of the ultimate stable states originate from the emergence history of flow elements, including the boundary layer, SWs, and compression waves. In Sec. \ref{sec:Numerical expriment}, 3D DNSs are performed to check the authenticity of the thought experiment.
\label{subsec:Conjecture}
\section {Numerical expriment}
\label{sec:Numerical expriment}
\par In this section, the thought experiment shown in Fig.\ref{fig:two_routes} is numerically instantiated using 3D DNSs. The flow field in step 1 of can be obtained with the classical Rankine-Hugoniot relations. Therefore, only four steps need to be simulated: step 2 and 3 of both Route I and II. Details of DNS are in Sec. \ref{subsec:Details_DNS}. Route I and II are instantiated in \ref{subsec:Step_3_Route_I} and \ref{subsec:Route_II}, respectively. Contrast of the bistable states is in  Sec. \ref{subsec:Contrast}.
\subsection{Details of DNSs for DCR flows} 
\label{subsec:Details_DNS}
\subsubsection{Governing equations and numerical methods}
\par The governing equations solved are dimensionless 3D Navier–Stokes equations for unsteady, compressible flow in curvilinear coordinates,
\begin{equation}
	\frac{\partial \mathbf{Q}}{\partial t}+\frac{\partial\left(\mathbf{F}_c+\mathbf{F}_v\right)}{\partial \xi}+\frac{\partial\left(\mathbf{G}_c+\mathbf{G}_v\right)}{\partial \eta}+\frac{\partial\left(\mathbf{H}_c+\mathbf{H}_v\right)}{\partial \zeta}=0,
\end{equation}
where $\mathbf{Q}$ is the conservative vector flux, $\mathbf{F}_c$, $\mathbf{G}_c$ and $\mathbf{H}_c$ are the invicid convection fluxes, $\mathbf{F}_v$, $\mathbf{G}_v$ and $\mathbf{H}_v$ are the viscous fluxes. Here, $\mathbf{Q}$, $\mathbf{F}_c$ and $\mathbf{F}_v$ are defined as 
\begin{equation}
	\mathbf{Q}=J\left(\begin{array}{c}
		\rho^* \\
		\rho^* \mathbf{U} \\
		\rho^* e^*
	\end{array}\right), \mathbf{F}_c=J\left(\begin{array}{c}
		\rho^* \mathbf{U}^{T} \mathbf{J}_{\xi} \\
		\rho^*  \mathbf{U}^{T} \mathbf{J}_{\xi} \mathbf{U} + p^* \mathbf{J}_{\xi} \\
		\left(\rho^*e^*+p^* \right) \mathbf{U}^{T} \mathbf{J}_{\xi}
	\end{array}\right), \mathbf{F}_v=J\left(\begin{array}{c}
		0 \\
		\frac{1}{\operatorname{Re}_{\infty}} \boldsymbol{\tau} \mathbf{J}_{\xi} \\
		\left(\frac{1}{\operatorname{Re}_{\infty}} \mathbf{U}^{T} \boldsymbol{\tau} - \frac{1}{(\gamma-1) M_{\infty}^2 \operatorname{Pr}} \mathbf{q^{T}}\right) \mathbf{J}_{\xi}
	\end{array}\right),
\end{equation}
where $\mathbf{J}_{\xi}=(\xi_x,\xi_y,\xi_z)^{T}$ are the metric coefficients and $J$ is the determinant of the Jacobian matrix transforming the Cartesian coordinates $(x,y,z)$ into the computational coordinates $(\xi,\eta,\zeta)$. The  velocity vector $\mathbf{U}$, the heat flux vector $\mathbf{q}$ and the stress tensor $\boldsymbol{\tau}$ are defined as
\begin{equation}
	\mathbf{U}=\left(\begin{array}{c}
		u^* \\
		v^* \\
		w^*
	\end{array}\right), \quad \mathbf{q}=\left(\begin{array}{c}
		q_{\xi}^* \\
		q_{\eta}^* \\
		q_{\zeta}^*
	\end{array}\right), \quad \boldsymbol{\tau}=\left(\begin{array}{c}
		\tau_{\xi \xi}^*, \tau_{\xi \eta}^*, \tau_{\xi \zeta}^* \\
		\tau_{\eta \xi}^*, \tau_{\eta \eta}^*, \tau_{\eta \zeta}^* \\
		\tau_{\zeta \xi}^*, \tau_{\zeta \eta}^*, \tau_{\zeta \zeta}^*
\end{array}\right).
\end{equation}
The total energy $e^*$ is defined as
\begin{equation}
e^*=\frac{u^{* 2}+v^{* 2}+w^{* 2}}{2}+\frac{1}{\gamma-1} \frac{p^*}{\rho^*}.
\end{equation}
The other four flux terms have the similar forms: $\mathbf{G}_c$ and $\mathbf{H}_c$ are similar in form to $\mathbf{F}_c$; $\mathbf{G}_v$ and $\mathbf{H}_v$ are similar in form to $\mathbf{F}_v$. The viscosity $\mu$ is determined with the Sutherland's law, and perfect gas equation, relating the pressure $p$ the density $\rho$ and the temperature $T$, is used to close the equations set. The Prandtl number $Pr = 0.7$ and the specific heat ratio $\gamma = 1.4$ are chosen in the simulations. The non-dimensional variable are normalized using the inflow parameters: $\rho^*=\rho / \rho_{\infty}$, $u^*=u / u_{\infty}$, $v^*=v / u_{\infty}$, $w^*=w / u_{\infty}$, $T^*=T / T_{\infty}$, $e^*= e / u_{\infty}^2$ and $p^*=\rho^* T^* / (\gamma Ma_{\infty}^2)$. The reference length is chosen as $1$ mm.
\par In terms of the numerical methods, the time integration is performed by the third-order TVD-type Runge–Kutta method; the inviscid fluxes are discreted by the fifth-order WENO method \citep{jiang1996efficient}; the viscid fluxes are discreted with the sixth-order central difference scheme. The DNSs are conducted with the in-house code OPENCFD-SC, whose capability has been well examined \citep{hu2017beta,xu2021effect}, \textcolor{blue}{especially in studying compression ramp flows.}
\subsubsection{Wall geometry, flow parameters, mesh spacing and initial \& boundary conditions}
\par
\begin{figure}
	\centering
	{\includegraphics[width = 0.7\columnwidth]{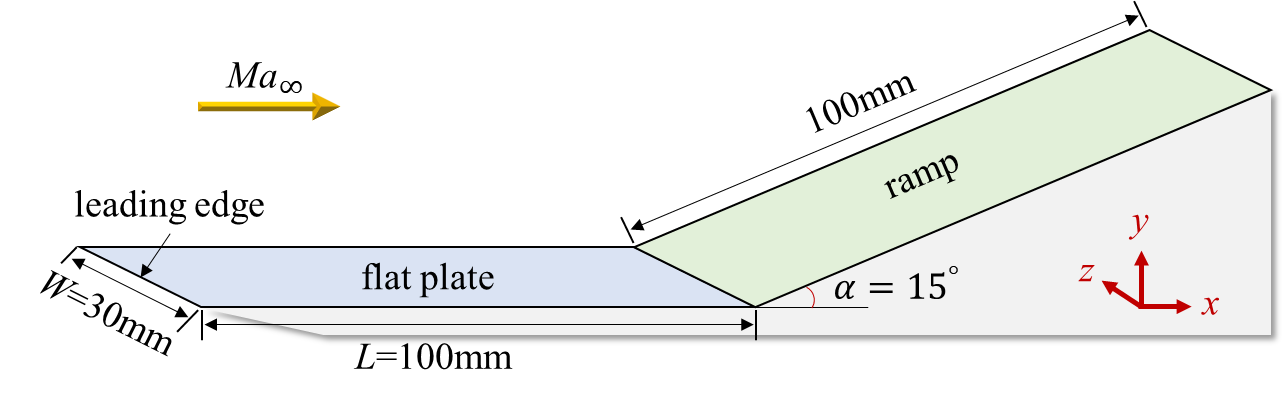}}
	\caption{The schematic of the DCR configuration.}
	\label{fig:wall_geometry}
\end{figure}

 The choices of the compression ramp geometry (DCR) and the flow parameters are based on the recent shock-tunnel experiments \citep{roghelia2017experimental,Roghelia2017ExperimentalIO,chuvakhov2017effect}. As the DCR configuration shown in Fig. \ref{fig:wall_geometry}, the DCR configuration is mainly composed of two parts, the flat plate and the ramp, both of which have the same length $L = 100$mm and width $W = 30$mm. The ramp angle is $\alpha = 15^{\circ}$. The flow parameters are shown in Tab. \ref{tab:Flow parameters}, including the Mach number $Ma_{\infty} =7.7$, the Reynolds number $Re_{\infty} = 4.2 \times 10^{6}$, the velocity $u_{\infty} = 1726 \text{m s}^{-1}$, the density $\rho_{\infty} = 0.021 \text{kg m}^{-3}$, the temperature $T_{\infty} = 125 \text{K}$ and the pressure $p_{\infty} = 760 \text{Pa}$. The isothermal wall condition is used with $T_{w} = 293 \text{K}$, since the run time of the shock tunnel is very short.
\begin{table}
	\begin{center}
		\def~{\hphantom{0}}
		\begin{tabular}{lcccccccc}
			&\quad $M_{\infty} \quad$ &\quad $Re_{\infty}$&\quad $h_{0}$ &\quad $u_{\infty}$ &\quad $\rho_{\infty}$ &\quad $T_{\infty}$ &\quad $p_{\infty}$ &\quad $T_{w}$ \\[3pt]
			&\quad   &\quad (mm$^{-1}$) &\quad (MJ kg$^{-1}$) &\quad (m s$^{-1}$) &\quad kg m$^{-3}$ &\quad (K) &\quad (Pa) &\quad (K) \\
			&\quad $7.7$ &\quad $4.2 \times 10^6$ &\quad $1.7$ &\quad $1726$ &\quad 0.021 &\quad $125$ &\quad $760$ &\quad $293$ 
		\end{tabular}
		\caption{The flow conditions based on the shock tunnel TH2 \citep{Roghelia2017ExperimentalIO}.}
		\label{tab:Flow parameters}
	\end{center}
\end{table}
\begin{figure}
	\centering
	\subfigure[\label{subfig:M1-mesh}]{\includegraphics[width = 0.45\columnwidth]{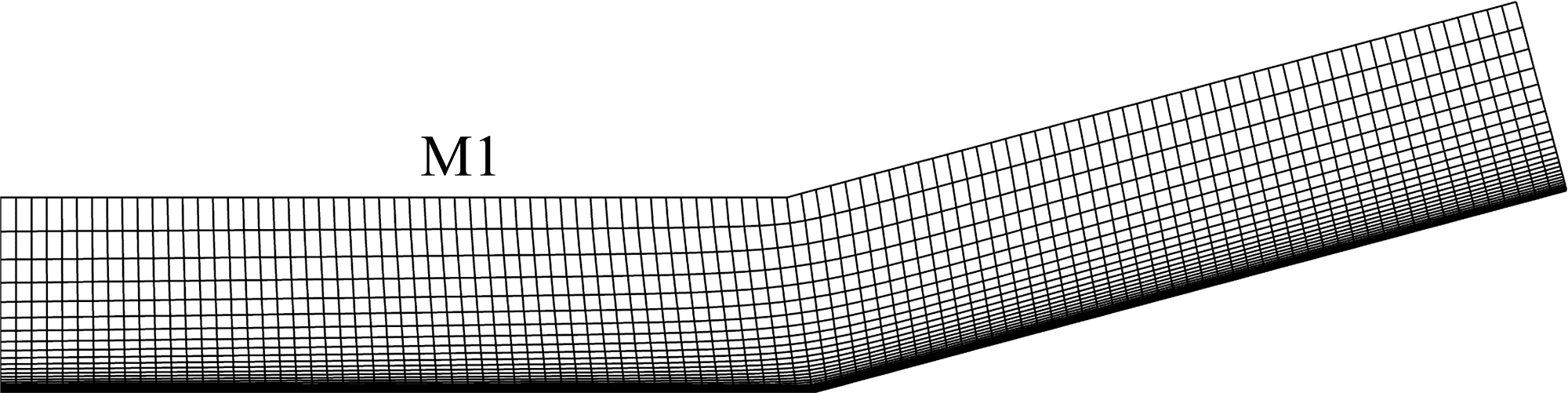}} \quad
	\subfigure[\label{subfig:M2-mesh}]{\includegraphics[width = 0.45\columnwidth]{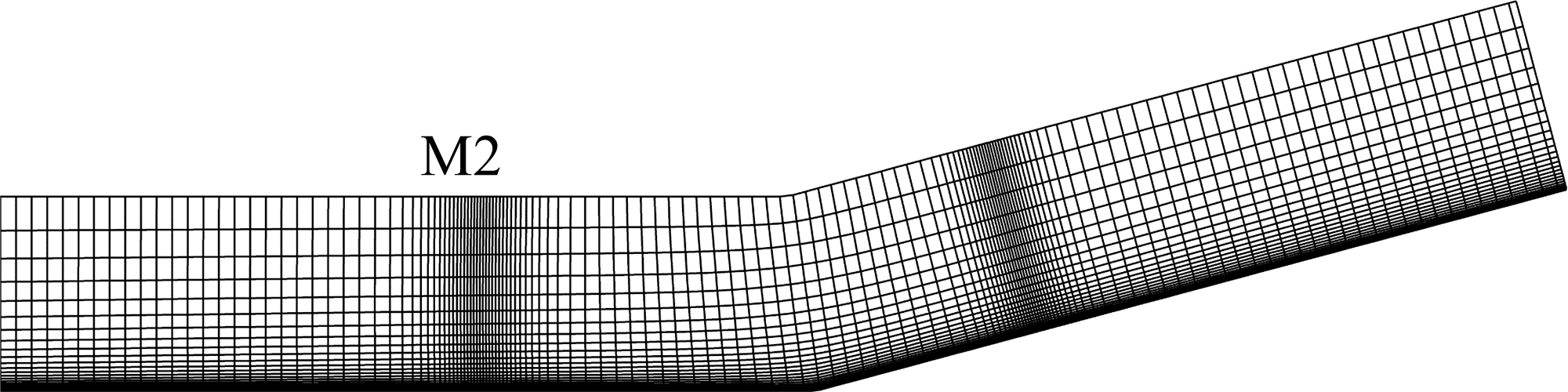}}
	\caption{Mesh distributions in $x-y$ plane. (a) M1; (b) M2.}
	\label{fig:Mesh_M1_M2}
\end{figure}
\par The computational domain, i.e., the mesh region, is shown in Fig. \ref{fig:Mesh_M1_M2}. The streamwise region is $-4 \text{mm} \le x \le 196.6 \text{mm}$ (the leading edge is at $x = 0$ mm); the wall-normal height is $25$ mm; the spanwise width is $30$ mm. Two mesh resolutions, $1020 \times 250 \times 200$ (M1) and $1420 \times 250 \times 300$ (M2), are considered for time and grid convergence studies. Both M1 and M2 cluster the grids near the wall with the first wall-normal grid height being fixed to $\Delta y_{\text{w}} = 0.008$ mm, yielding the non-dimensional height $\Delta y_{\text{w}}^{+} \approx 0.3$ mm at $x/L \approx 0.5$. Spanwise grid spaces of both M1 and M2 are uniform, as $\Delta z = 0.15$ mm (M1) and $0.1$ mm (M2). The streamwise grid spaces are uniform as $\Delta x_{\xi} = 0.2$ mm in M1. To investigate the pressure gradient effects, near the separation point $x_{S}/L \approx 0.55$ and the reattachment point $x_{R}/L \approx 1.28$ (also shown in Fig. \ref{fig:DNS_M1_M2}) in M2, the $\Delta x_{\xi}$ spaces are severally clustered with $200$ points in the streamwise direction.
\par For the initial conditions, the initial flow fields for M1 and M2 are both \textcolor{blue}{extruded (or extended)} spanwisely from the 2D simulation result (denoted as D$^{\text{I}}_{0}$), in which field the separation point $x_{S}/L \approx 0.55$ and the reattachment point $x_{R}/L \approx 1.28$. For the boundary conditions, the free stream is set both at the numerical inlet boundary ($x = -4$mm) and the upper boundary; no-slip and isothermal ($T_w = 293$K) conditions are set for $x \ge 0$mm on the wall; the \textcolor{blue}{extrapolation} condition is set at the outflow boundary.
\subsubsection{Verification and validation}
\begin{figure}
	\centering
	\subfigure[\label{subfig:Numerical_Process1-step2_M1}]{\includegraphics[width = 0.48\columnwidth]{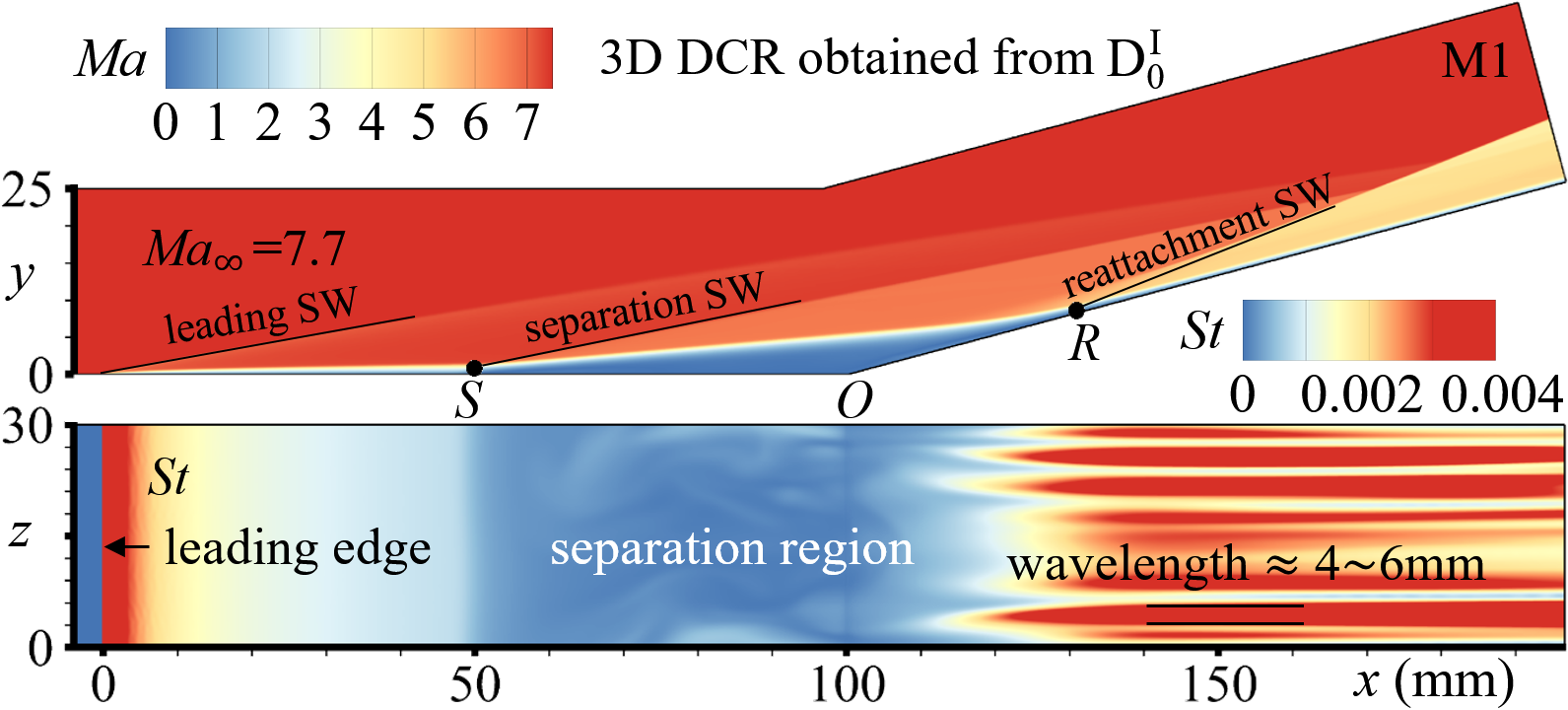}}
	\subfigure[\label{subfig:Numerical_Process1-step2_M2}]{\includegraphics[width = 0.48\columnwidth]{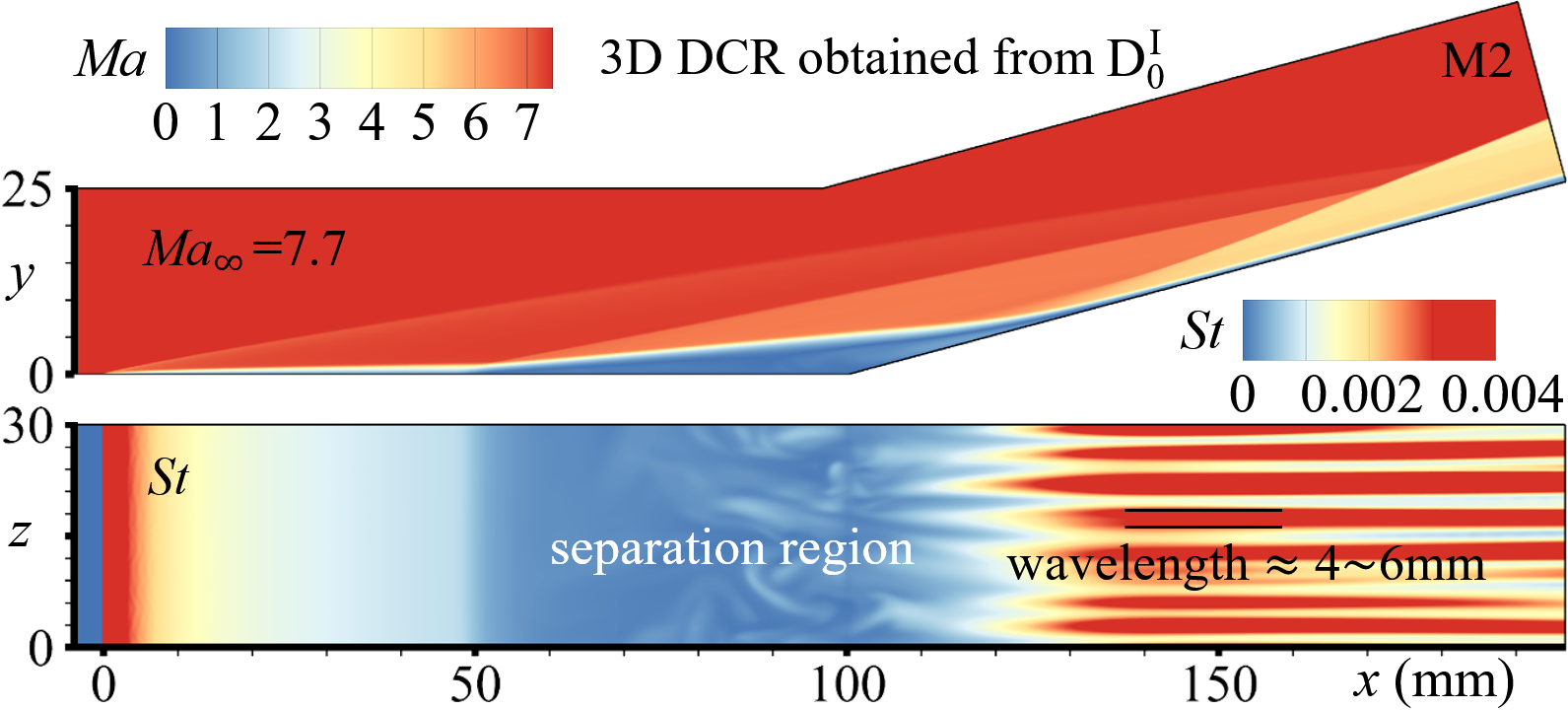}}
	\caption{DNS results colored by local Mach number in $x-y$ plane and the Stanton number $St$ in $x-z$ plane. (a) M1; (b) M2.}
	\label{fig:DNS_M1_M2}
\end{figure}
\par The 3D instantaneous flow fields of M1 and M2 are shown in Fig. \ref{fig:DNS_M1_M2}. In the central $x-y$ planes, it clearly shows that the SW configurations include the leading edge SW, the separation SW, and the reattachment SW. In the first $x-z$ planes (on the wall), it can be noted that the wavelengths of the spanwise streaks on the ramp, colored by the Stanton number $St$ defined in Eq. \ref{eq:Cf_Cp_St}, are about $4 \sim 6$mm, which are consistent with the previous experimental observations \citep{roghelia2017experimental,Roghelia2017ExperimentalIO,chuvakhov2017effect} and numerical results \citep{cao2021unsteady}.
\begin{figure}
	\centering
	\subfigure[\label{subfig:TimeConvergence}]{\includegraphics[width = 0.48\columnwidth]{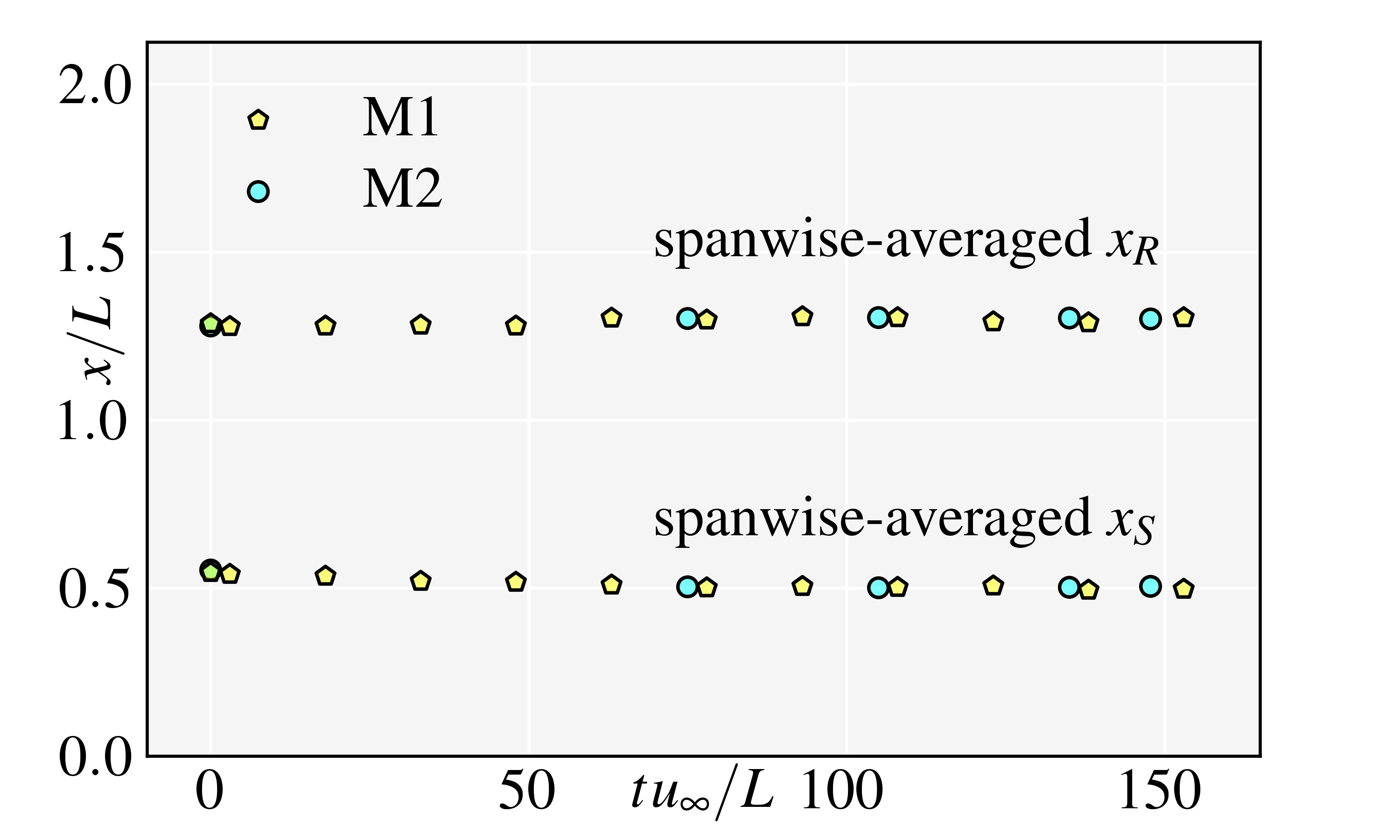}}
	\subfigure[\label{subfig:Validation_convergence_Cf}]{\includegraphics[width = 0.48\columnwidth]{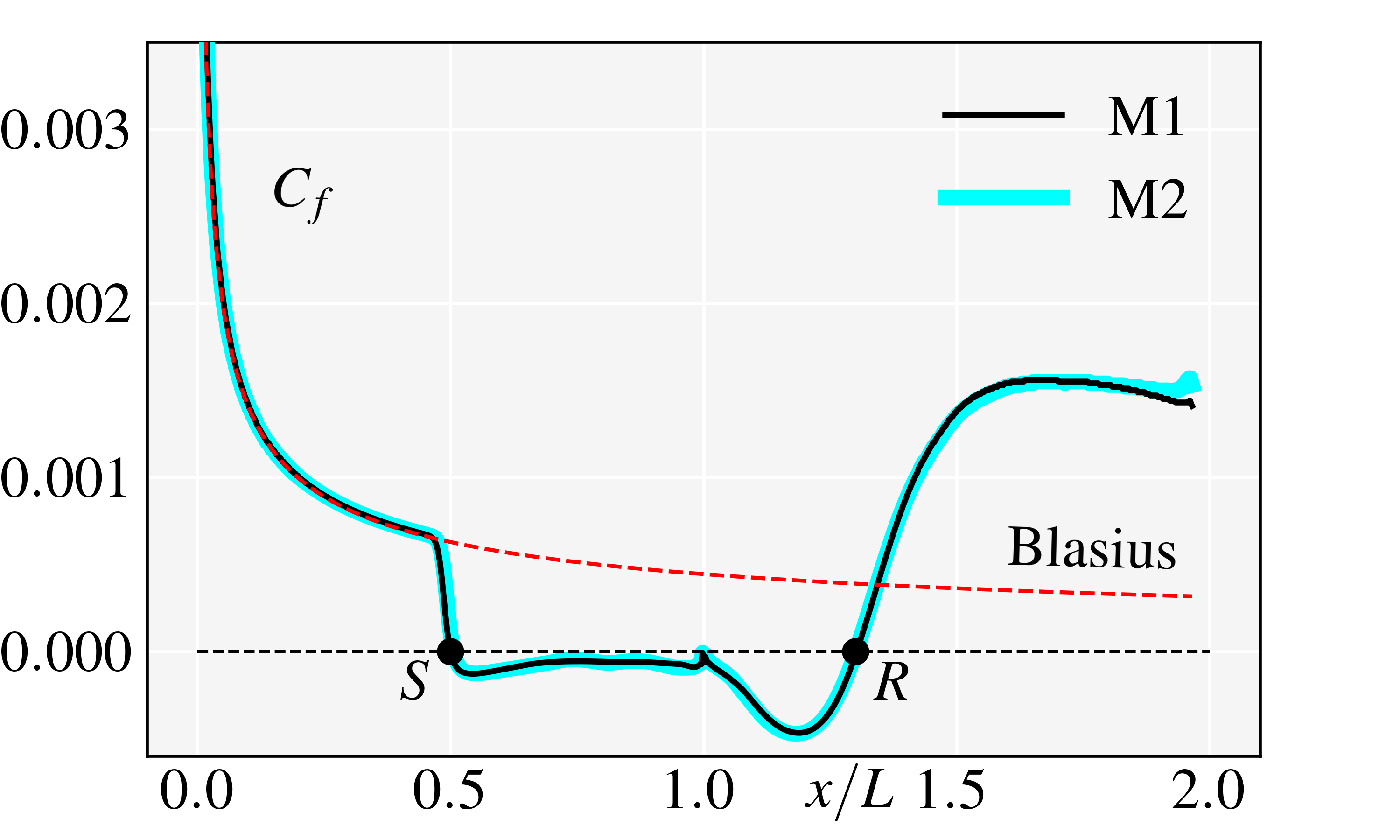}}
	\subfigure[\label{subfig:validation_Blasius}]{\includegraphics[width = 0.48\columnwidth]{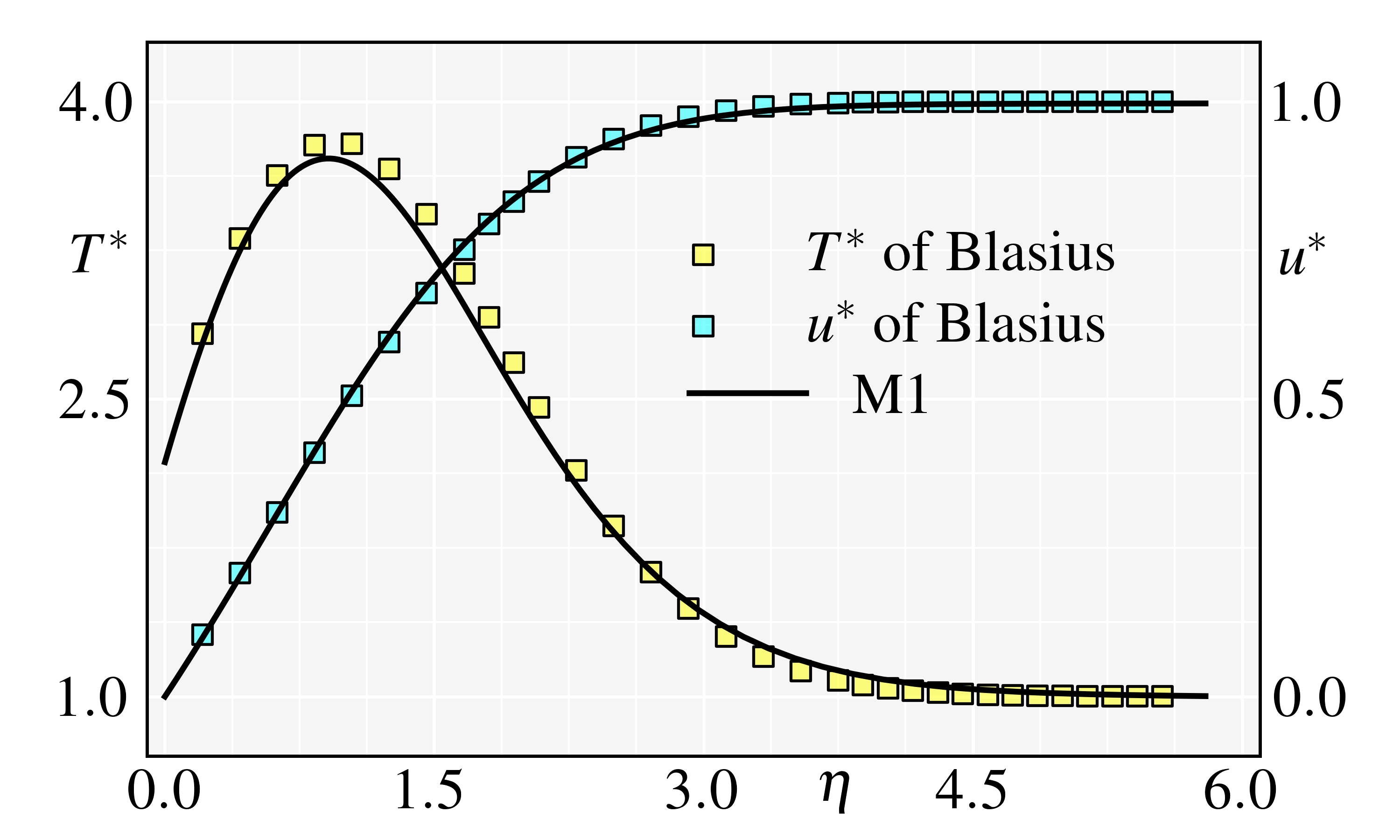}}
	\subfigure[\label{subfig:Validation_Cp_St}]{\includegraphics[width = 0.48\columnwidth]{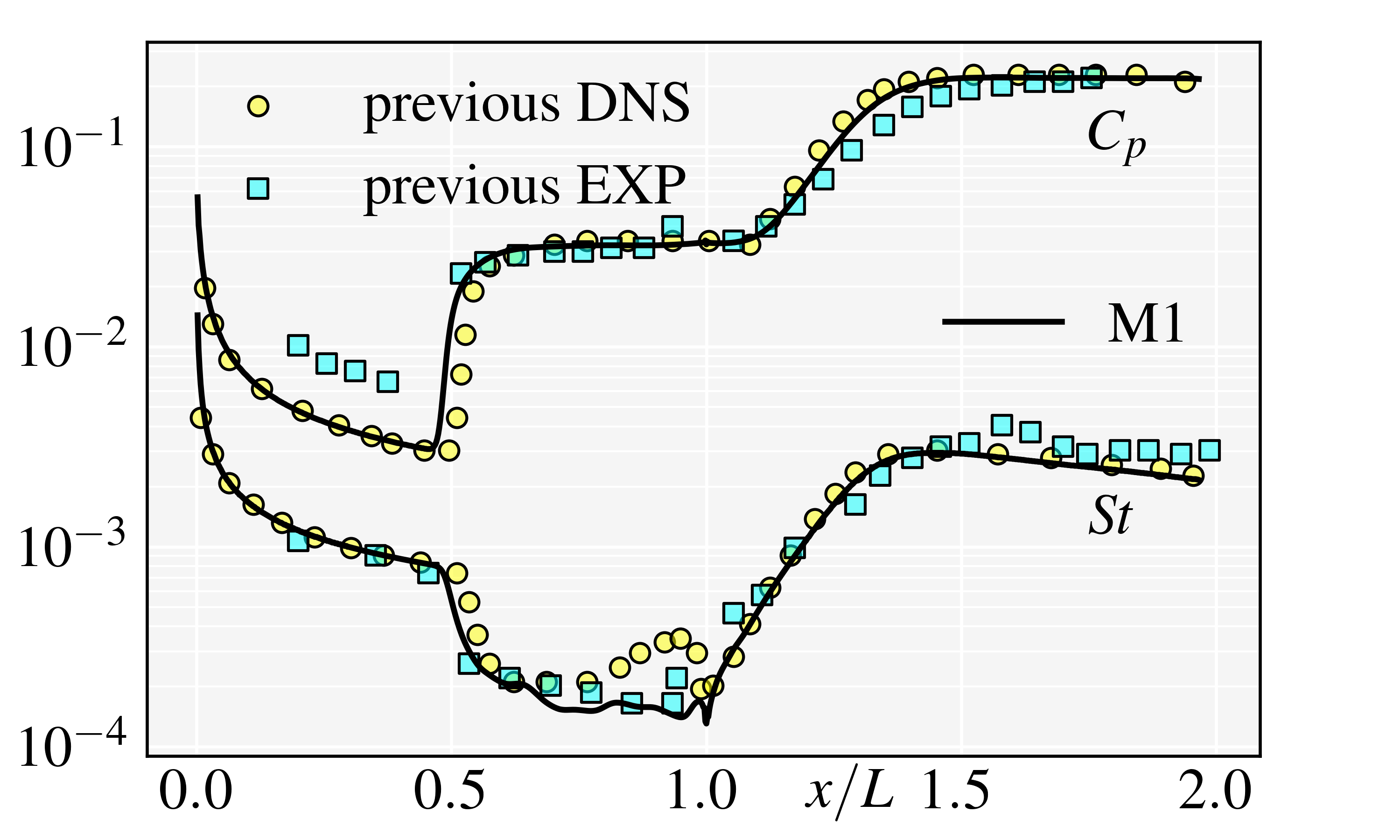}}
	\vspace{-2.5mm}
	\caption{(a) Verifications of the time-convergence with variations of $x_S$ and $x_R$; (b) verification of the grid-convergence with $C_f$ distributions, and validation with $C_f$ on the flat-plate region; (c) validation with theoretical normal profiles of $u$ and $T$; (d) validation with $C_p$ and $St$ distributions of previous studies.}
\end{figure}
The aerothermodynamic characteristics are used to verify and validate the DNSs' results, including the skin friction coefficient $C_{f}$, the surface pressure coefficient $C_{p}$, and the Stanton number $St$, which are defined as
\begin{equation}
	C_f=\frac{\tau_{w}}{\frac{1}{2}\rho_{\infty} u_{\infty}^2}, \quad C_p=\frac{p_{w} - p_{\infty}}{\frac{1}{2} \gamma Ma_{\infty}^2 p_{\infty}}, \quad St = \frac{q_w}{\rho_{\infty} u_{\infty} c_{p}(T_{aw} - T_{w})}
\label{eq:Cf_Cp_St}
\end{equation}
where $\tau_{w}$, $p_{w}$ and $q_{w}$ are the friction, pressure and heat flux on the wall, respectively. $c_p$ and $T_{aw}$ are the specific heat capacity and the adiabatic wall temperature, respectively. 
\par The verifications include the time- and grid-convergences. For the time-convergences, as shown in Fig. \ref{subfig:TimeConvergence}, the non-dimensional simulation time $tu_{\infty}/L$ of M1 and M2 are both longer than $145$. It shows that the spanwise-averaged locations of the separation and reattachment points only oscillate slightly near $x_{S}/L \approx 0.5$ and $x_{R}/L \approx 1.3$ after $tu_{\infty}/L > 70$, implying the separated DCR flows have been stabilized in both M1 and M2. For the grid-convergence, as shown in Fig. \ref{subfig:Validation_convergence_Cf}, there are only small discrepancy of the $C_f$ distributions between M1 and M2, implying the mesh resolution of M1 is suitable for the present simulations.
\par The validations include the comparisons of the present $C_f$, $C_p$, $S_t$distributions, and $u^*$, $T^*$ profiles with the \textcolor{blue}{accepted} theoretical, numerical and experimental results. The laminar boundary layer before separation is self-similar satisfying the Blasius theoretical solutions \citep{white2006viscous}. Fig. \ref{subfig:Validation_convergence_Cf} shows the theoretical $C_f$ distribution, and Fig. \ref{subfig:validation_Blasius} shows the theoretical normal profiles of $u^*$ and $T^*$ at $x/L = 0.36$. It is noted that both the present $C_f$ distribution and normal profiles agree well with the theoretical solutions, validating the simulations' accuracy. Furthermore, as shown in Fig. \ref{subfig:Validation_Cp_St}, the present $C_p$ and $St$ distributions are also compared with the previous experiment \citep{roghelia2017experimental,chuvakhov2017effect} and DNS \citep{cao2021unsteady} results, and the excellent agreements validate the present DNSs.
\subsection{DNSs for bistable states of CCR flows}
\subsubsection {Numerical strategy to realize the thought experiment}
\begin{figure}
	\centering
	{\includegraphics[width = 0.7\columnwidth]{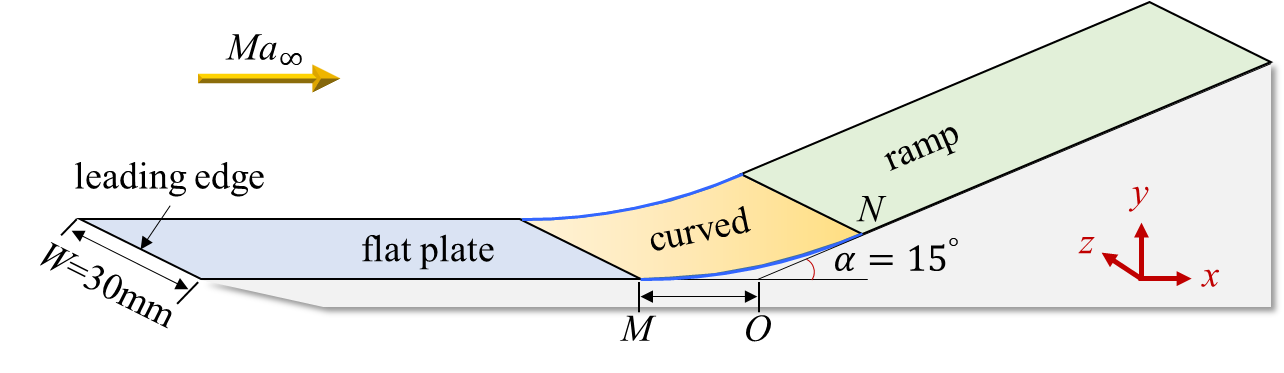}}
	\caption{The schematic of the DCR configuration.}
	\label{fig:curved_wall_geometry}
\end{figure}
\par As mentioned in Sec. \ref{sec:Thought experiment}, ``adding the viscosity'' and ``filling the corner'' are the key operations to realize the thought experiment. In terms of ``adding the viscosity'', the \textcolor{blue}{implement} method is changing the governing equations from Euler to Navier-Stokes. In terms of ``filling the corner'', since it requires to fill the corner macroscopical continuously (fill the corner atom by atom, microscopically), the strict operation is to manipulate the flow fields with an enormous number of steps, the number of which is the order of magnitude of the Avogadro number. Obviously, it is impossible in numerical simulation, and the equivalent implement is to replace the filling mode from `macroscopically continuous' to `macroscopically discrete'. The specific strategy is as follows.
\par \textit{For Route I} (``adding the viscosity'' $\rightarrow$ ``filling the corner''), the simulation process can be denoted as D$^{\text{I}}_{0}$ $\rightarrow$ C$_{1}^{\text{I}}$ $\rightarrow$ C$_{2}^{\text{I}}$ $\rightarrow$ ... $\rightarrow$ C$_{\text{N}}^{\text{I}}$, where the superscript `I' denotes Route I, and the subscript `N' represents the discrete number. Thus, D$^{\text{I}}_{0}$ is the stable separated DCR flow in step 2, and C$_{\text{N}}^{\text{I}}$ is the stable CCR flow after `N' times of corner filling in step 3.
\par \textit{For Route II} (``filling the corner'' $\rightarrow$ ``adding the viscosity''), the simulation process is denoted as C$_{\text{N}}^{\text{II,inv}}$ $\rightarrow$ C$_{\text{N}}^{\text{II}}$ ,where C$_{\text{N}}^{\text{II,inv}}$ is the inviscid CCR flow in step 2, and C$_{\text{N}}^{\text{II}}$ is the stale viscous flow in step 3, with the expectation of maintaining attached.
\begin{figure}
	\centering
	\subfigure[\label{subfig:CCR23_time}]{\includegraphics[width = 0.48\columnwidth]{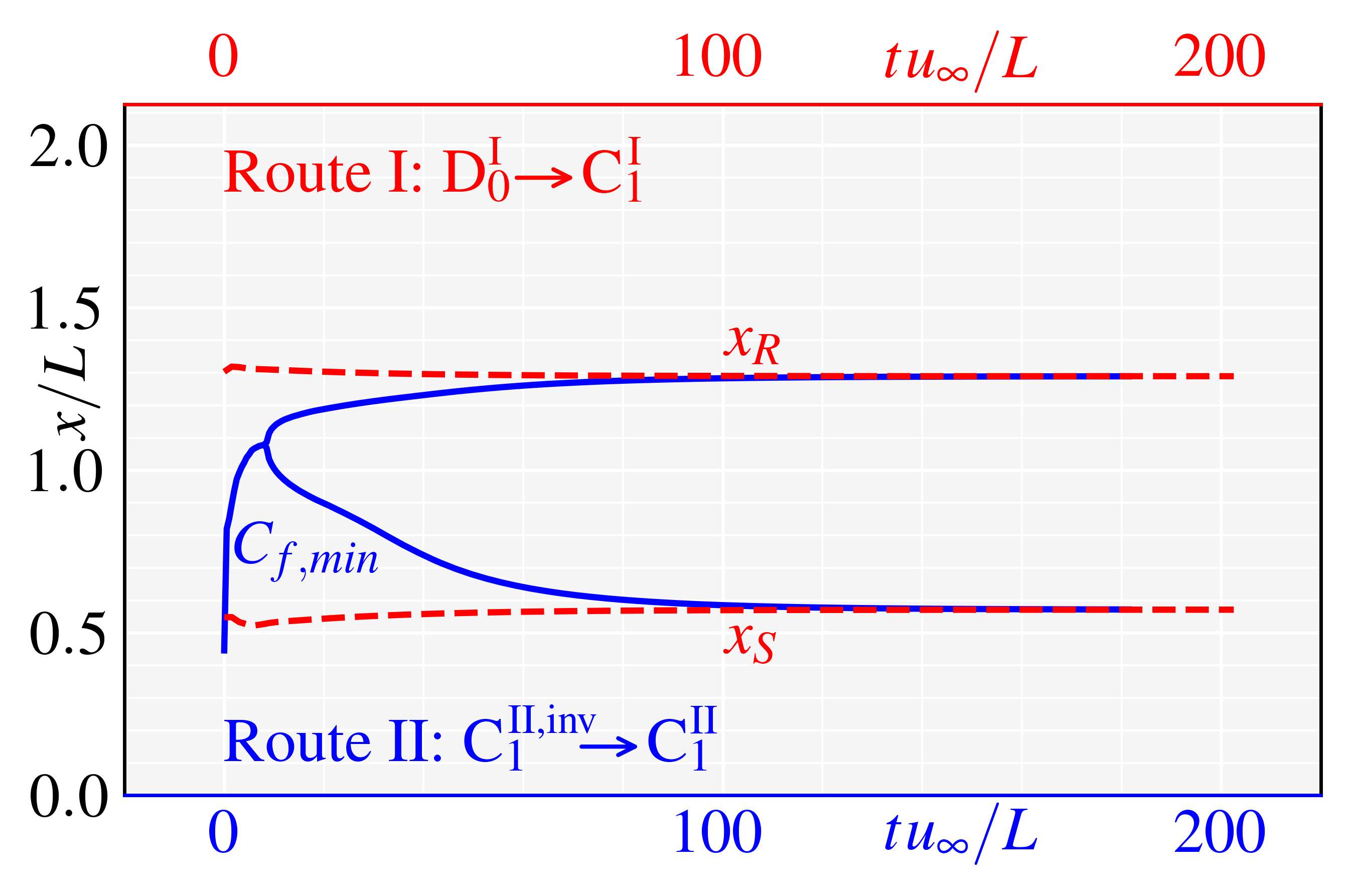}}
	\subfigure[\label{subfig:CCR25_time}]{\includegraphics[width = 0.48\columnwidth]{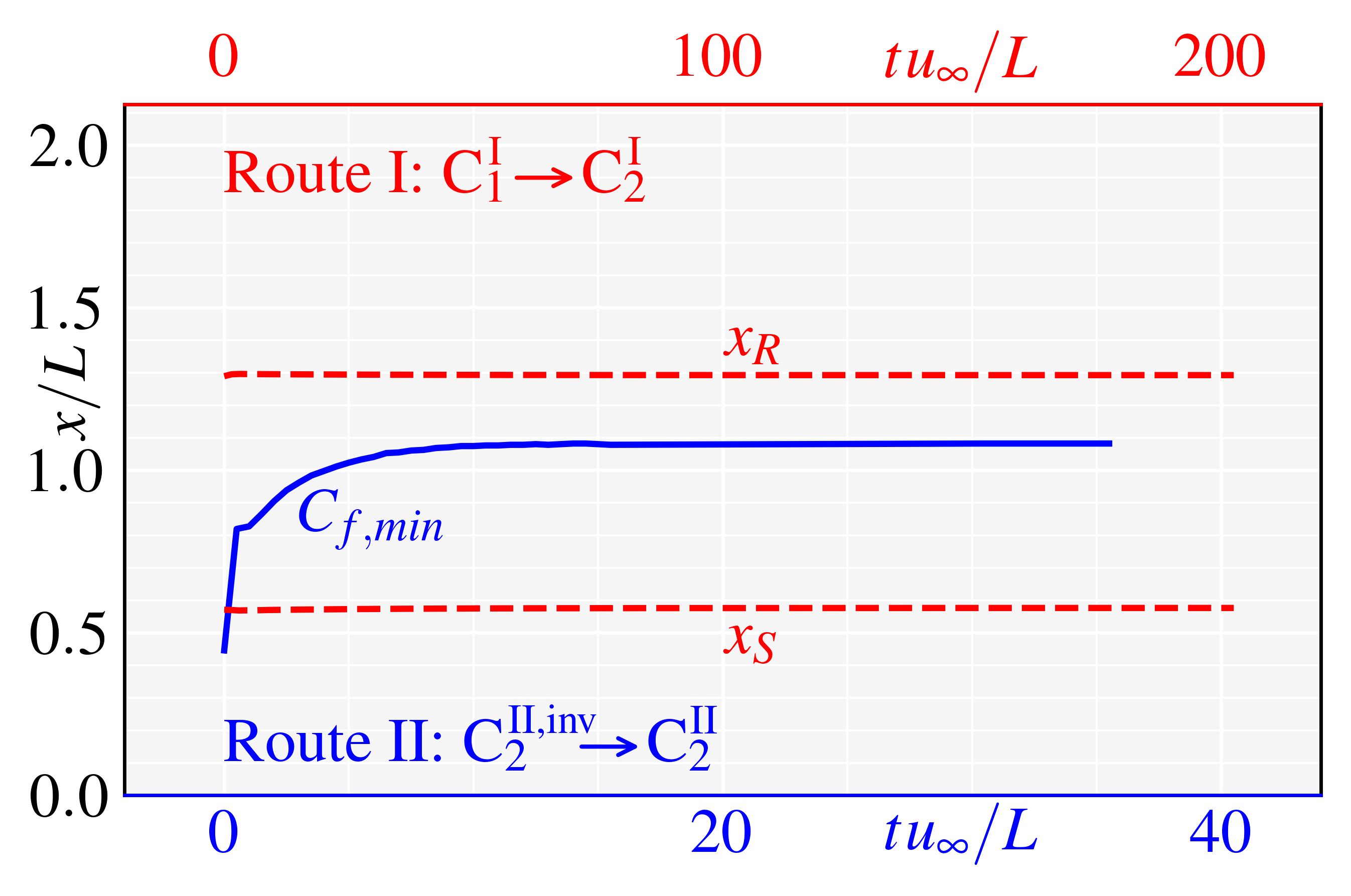}}
	\subfigure[\label{subfig:CCR29_time}]{\includegraphics[width = 0.48\columnwidth]{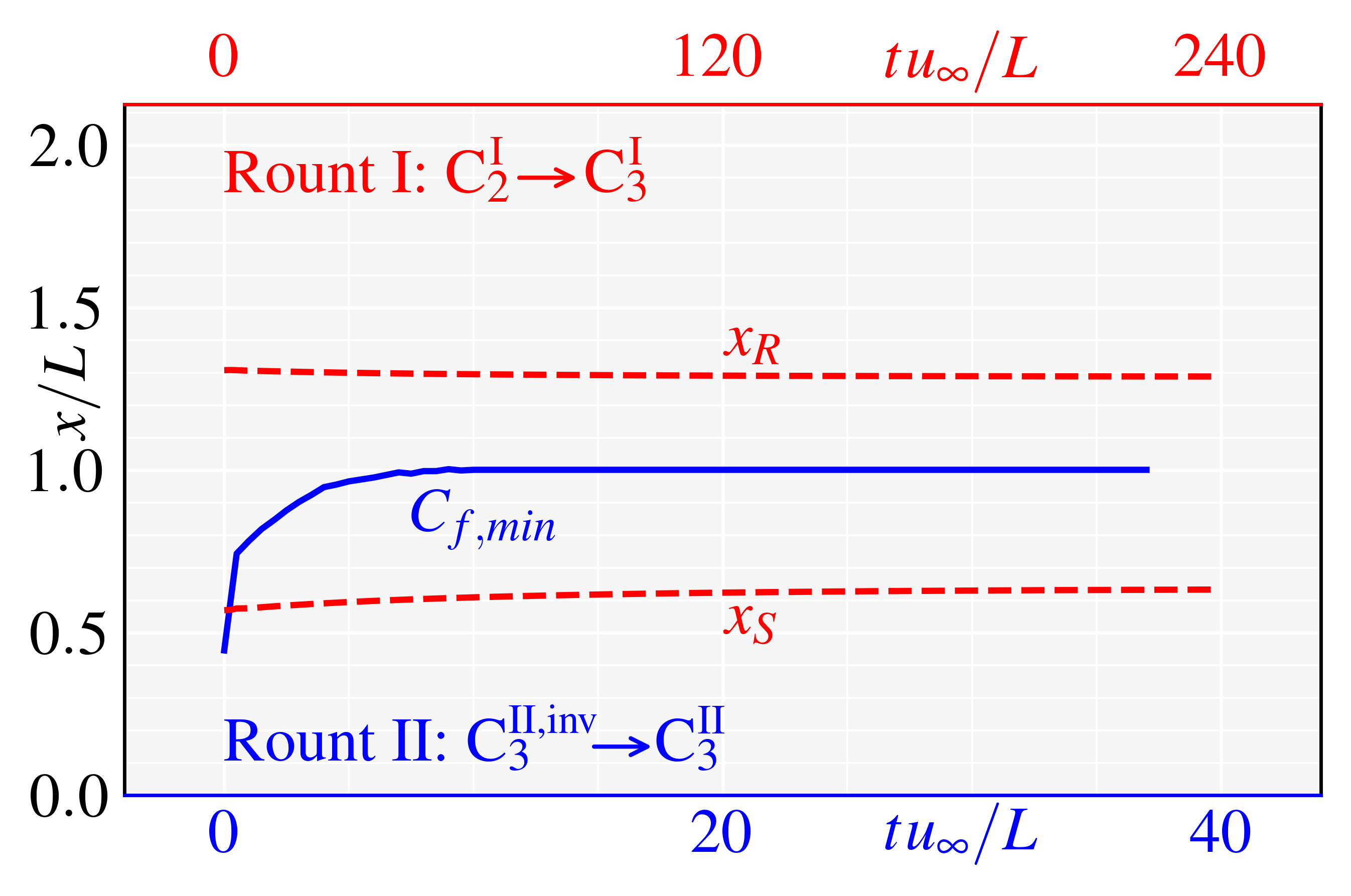}}
	\subfigure[\label{subfig:CCR30_time}]{\includegraphics[width = 0.48\columnwidth]{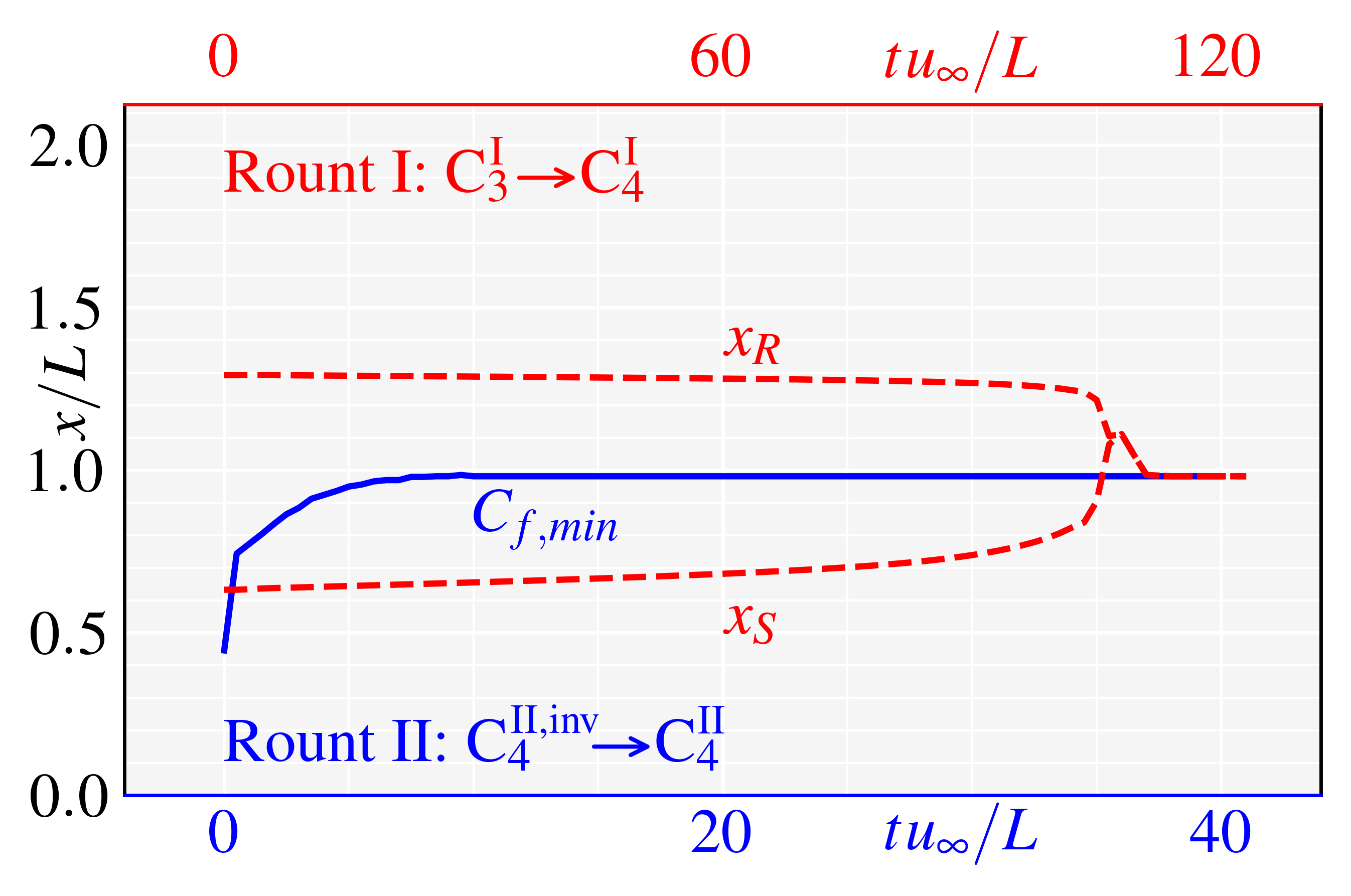}}
	\vspace{-2.5mm}
	\caption{Convergence trajectories of $x_{S}$, $x_{R}$, and $x_{C_{f,min}}$ of Route I and II. (a) C$_{1}^{\text{I}}$ and C$_{\text{1}}^{\text{II}}$; (b) C$_{2}^{\text{I}}$ and C$_{\text{2}}^{\text{II}}$; (c) C$_{3}^{\text{I}}$ and C$_{\text{3}}^{\text{II}}$; (d) C$_{4}^{\text{I}}$ and C$_{\text{4}}^{\text{II}}$.}\label{fig:Bistable_Convergence}
\end{figure}
\subsubsection {Simulations of Route I}
\label{subsec:Step_3_Route_I}
\par For D$^{\text{I}}_{0}$ $\rightarrow$ C$_{1}^{\text{I}}$ $\rightarrow$ C$_{2}^{\text{I}}$ $\rightarrow$ ... $\rightarrow$ C$_{\text{N}}^{\text{I}}$ (Route I), we set N $=$ 4, and C$_{1}$, C$_{2}$, C$_{3}$, C$_{4}$ are four CCR configurations with curvature radiuses $R \approx 174.7, 189.9, 220.3, 227.9$mm ($MO = 23, 25, 29, 30$mm), respectively. The four configurations' meshes are set with the same resolution of M1 ($1020 \times 250 \times 200$). Three characteristic locations, $x_{S}$, $x_{R}$, and $x_{C_{f,min}}$, are considered to quantify the flow states variations, where $x_{C_{f,min}}$ denotes the locations of minimal $C_{f}$ values in attached states. During process D$^{\text{I}}_{0}$ $\rightarrow$ C$_{1}^{\text{I}}$ $\rightarrow$ C$_{2}^{\text{I}}$ $\rightarrow$ C$_{\text{3}}^{\text{I}}$ $\rightarrow$ C$_{\text{4}}^{\text{I}}$, the evolutionary histories of spanwise-averaged $x_{S}$ and $x_{R}$ (red dashed lines) in C$_{1}$, C$_{1}$, C$_{3}$, and C$_{4}$ are shown in Fig.\ref{subfig:CCR23_time}, \ref{subfig:CCR25_time}, \ref{subfig:CCR29_time}, and \ref{subfig:CCR30_time}, respectively, and the non-dimensional time $tu_{\infty}/L > 100$ (see the upper red $x$-axes) for all cases. Obviously, $x_{S}$ and $x_{R}$ in C$_{1}$, C$_{2}$, and C$_{3}$ are all maintaining two branches for $tu_{\infty}/L > 200$, implying C$_{1}^{\text{I}}$, C$_{2}^{\text{I}}$, and C$_{\text{3}}^{\text{I}}$ can all be stabilized at separated states. However, for C$_{4}$ (the largest curvature radius), $x_{S}$ and $x_{R}$ ultimately merge into one branch (the $x_{C_{f,min}}$ branch— the blue solid line) at $tu_{\infty}/L \approx 100$, implying C$_{\text{4}}^{\text{I}}$ ends up in the attached state. Flow fields of C$_{1}^{\text{I}}$, C$_{2}^{\text{I}}$, C$_{\text{3}}^{\text{I}}$, and C$_{\text{4}}^{\text{I}}$ are also shown in Fig. \ref{subfig:C1_SS}, \ref{subfig:C2_SS}, \ref{subfig:C3_SS}, and \ref{subfig:C4_SA}, respectively. It is noted that, on the ramp, the streaks in CCR are weaker than those in DCR.
\subsubsection {Simulations of Route II}
\label{subsec:Route_II}
\begin{figure}
	\centering
	\subfigure[\label{subfig:CCR_inv-step2}]{\includegraphics[width = 0.48\columnwidth]{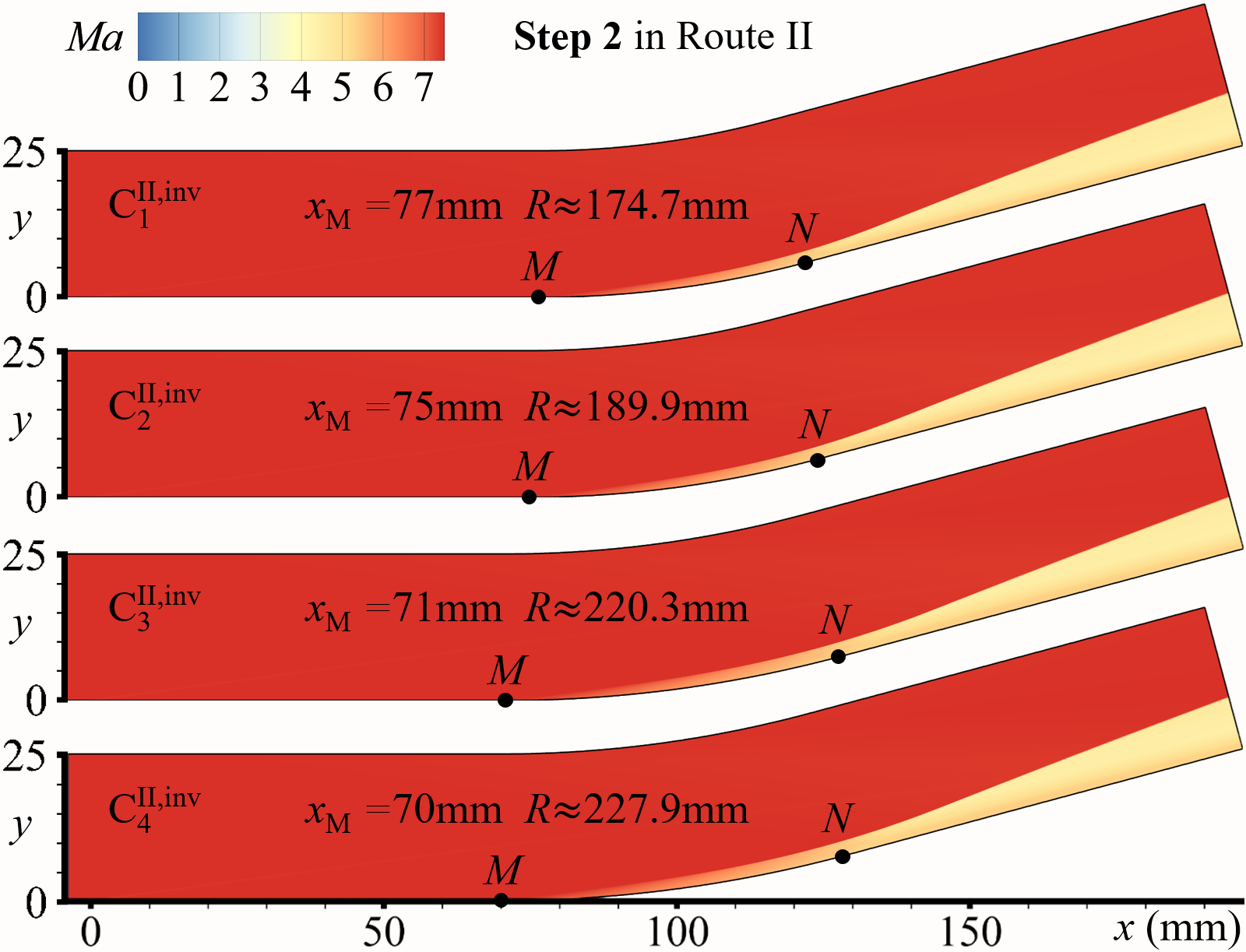}} 
	\subfigure[\label{subfig:CCR_Cp_inv-step2}]{\includegraphics[width = 0.48\columnwidth]{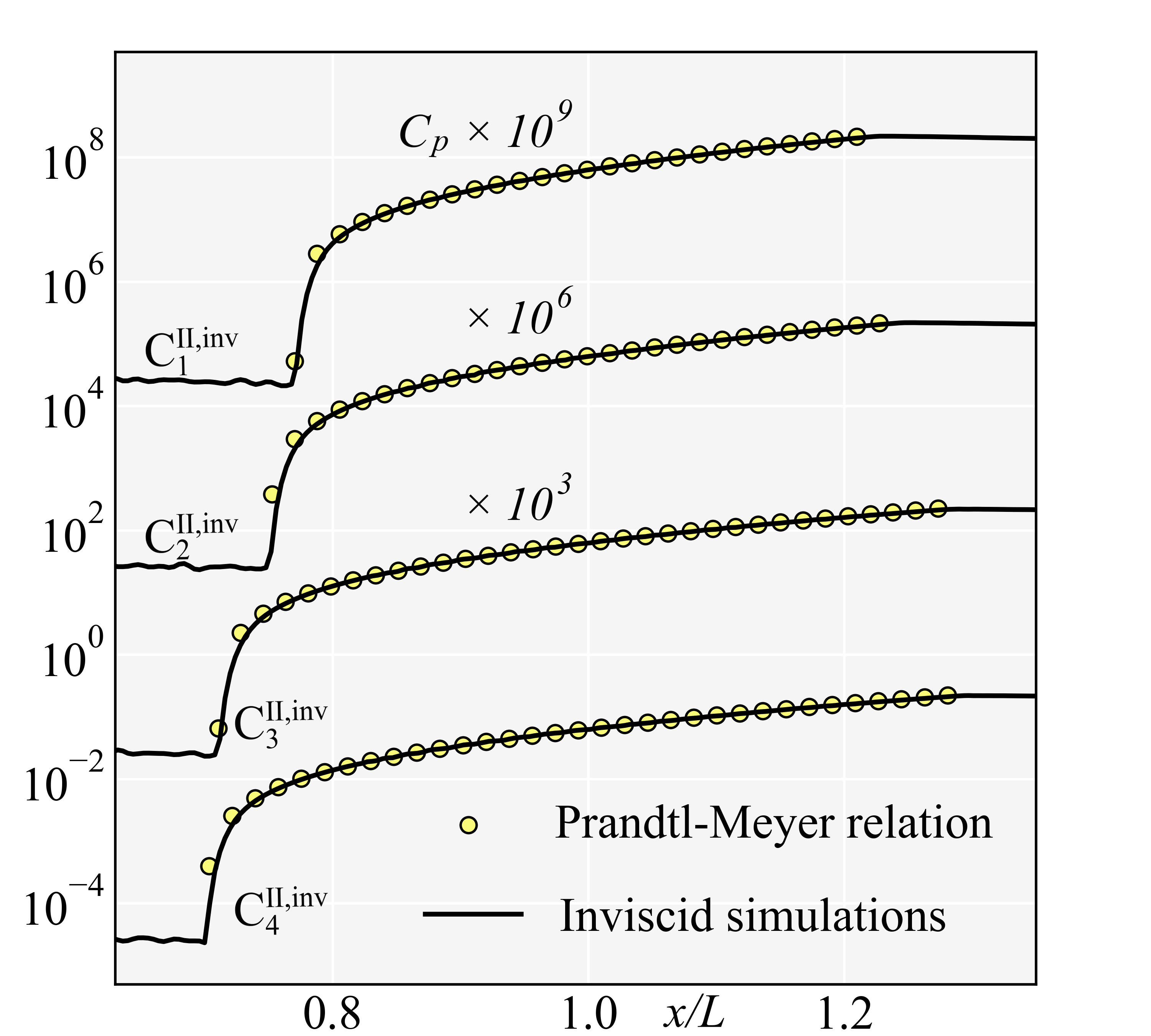}}
	\caption{Simulations of C$_{\text{1}}^{\text{II,inv}}$, C$_{\text{2}}^{\text{II,inv}}$, C$_{\text{3}}^{\text{II,inv}}$, and C$_{\text{4}}^{\text{II,inv}}$ to reproduce Step 2 in Route II. (a) flow fields colored by local Mach number in $x-y$ plane; (b) validation with Prandtl-Meyer relation of $C_p$.}\label{fig:DNS_inv_CCR}
\end{figure}
\par For C$_{\text{N}}^{\text{II,inv}}$ $\rightarrow$ C$_{\text{N}}^{\text{II}}$ (Route II), the flows in Step 2 (inviscid flows in C$_{\text{N}}$, N=4) are simulated with Euler equations, which are shown in Fig. \ref{subfig:CCR_inv-step2} and denoted as C$_{\text{1}}^{\text{II,inv}}$, C$_{\text{2}}^{\text{II,inv}}$, C$_{\text{3}}^{\text{II,inv}}$, and C$_{\text{4}}^{\text{II,inv}}$, respectively. A series of compression waves distribute on the curved wall $\overline{MN}$ regions, and the $C_p$ distributions on $\overline{MN}$ satisfy the Prandtl-Meyer relations, as shown in Fig. \ref{subfig:CCR_Cp_inv-step2},  validating the present simulations. 
\par
\begin{figure}
	\centering
	\subfigure[\label{subfig:C1_SS}]{\includegraphics[width = 0.48\columnwidth]{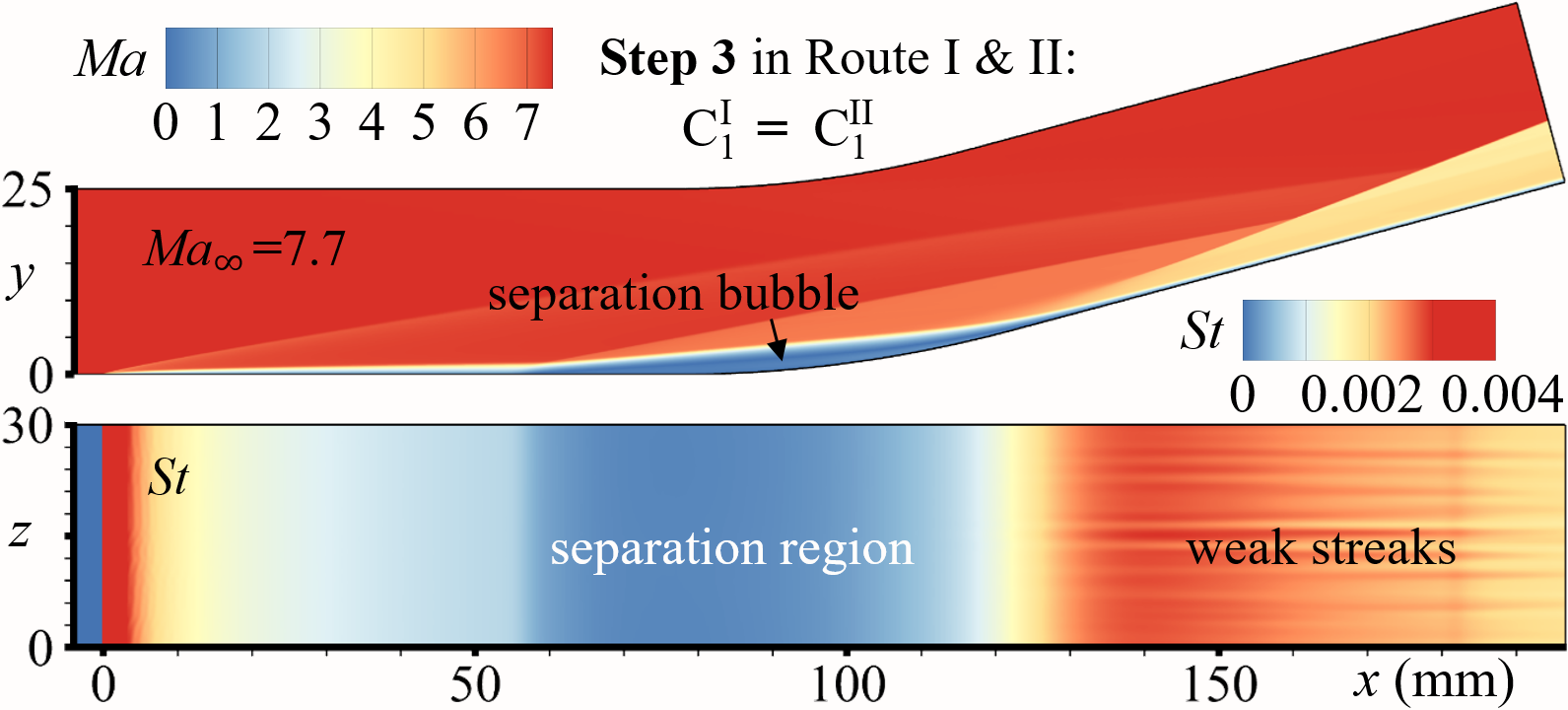}} 
	\subfigure[\label{subfig:C4_SA}]{\includegraphics[width = 0.48\columnwidth]{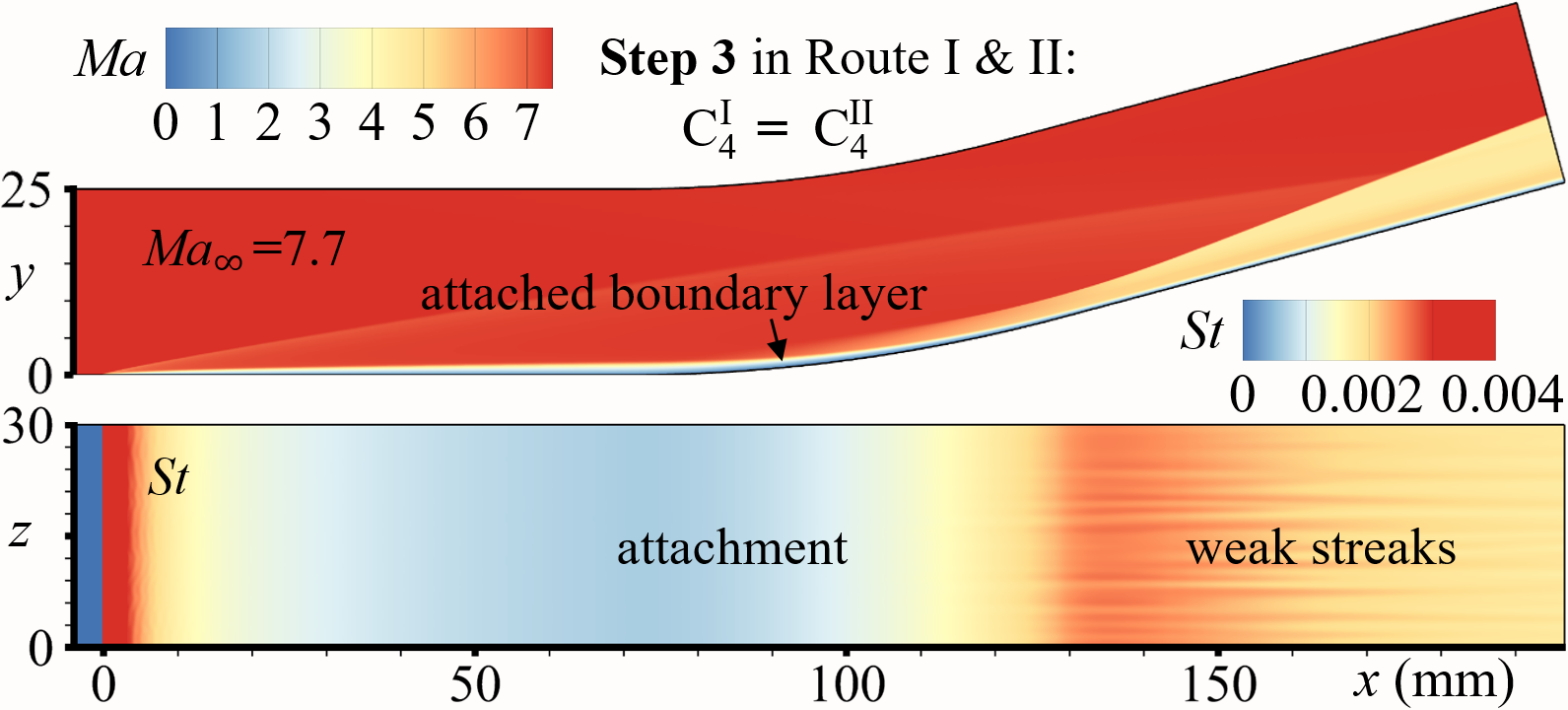}} 
	\subfigure[\label{subfig:C2_SS}]{\includegraphics[width = 0.48\columnwidth]{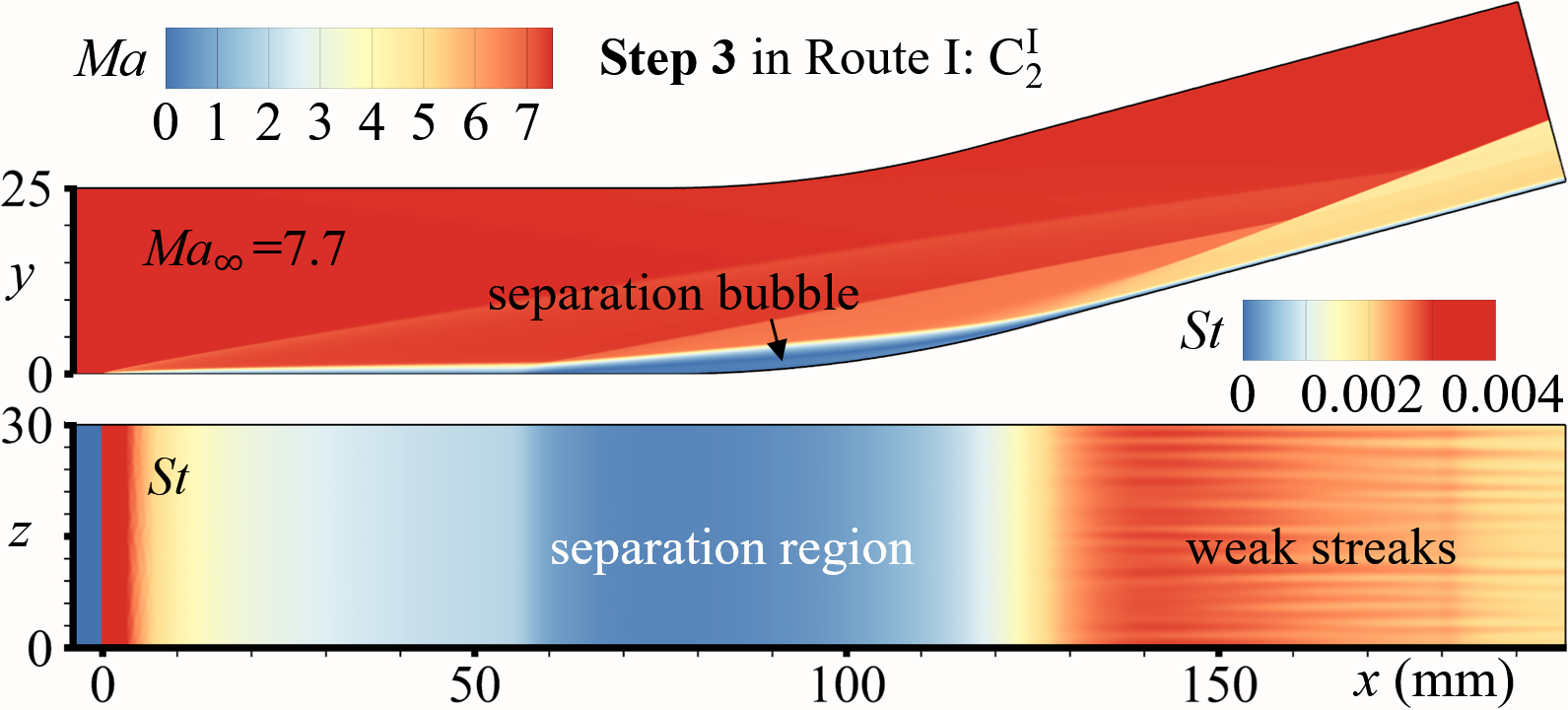}} 
	\subfigure[\label{subfig:C2_SA}]{\includegraphics[width = 0.48\columnwidth]{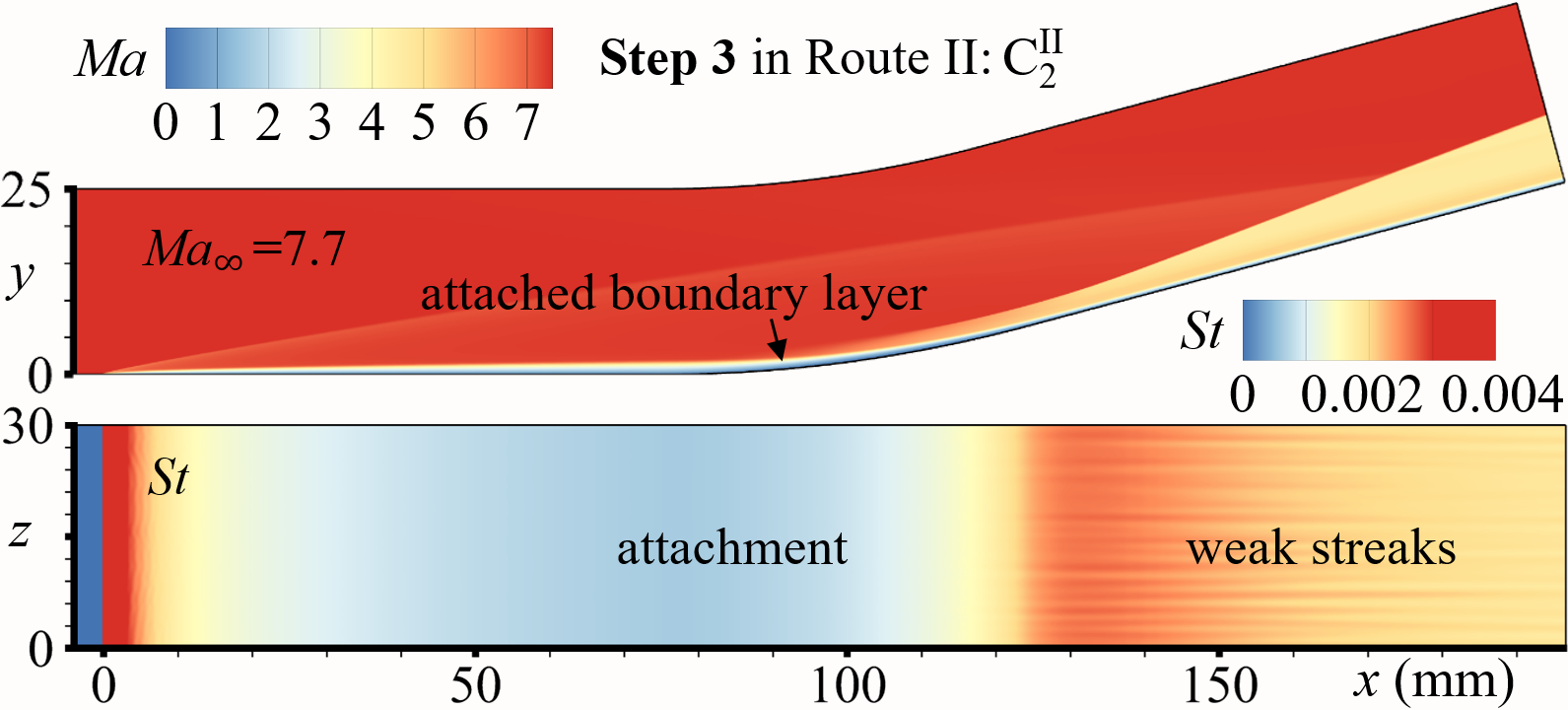}} 
	\subfigure[\label{subfig:C3_SS}]{\includegraphics[width = 0.48\columnwidth]{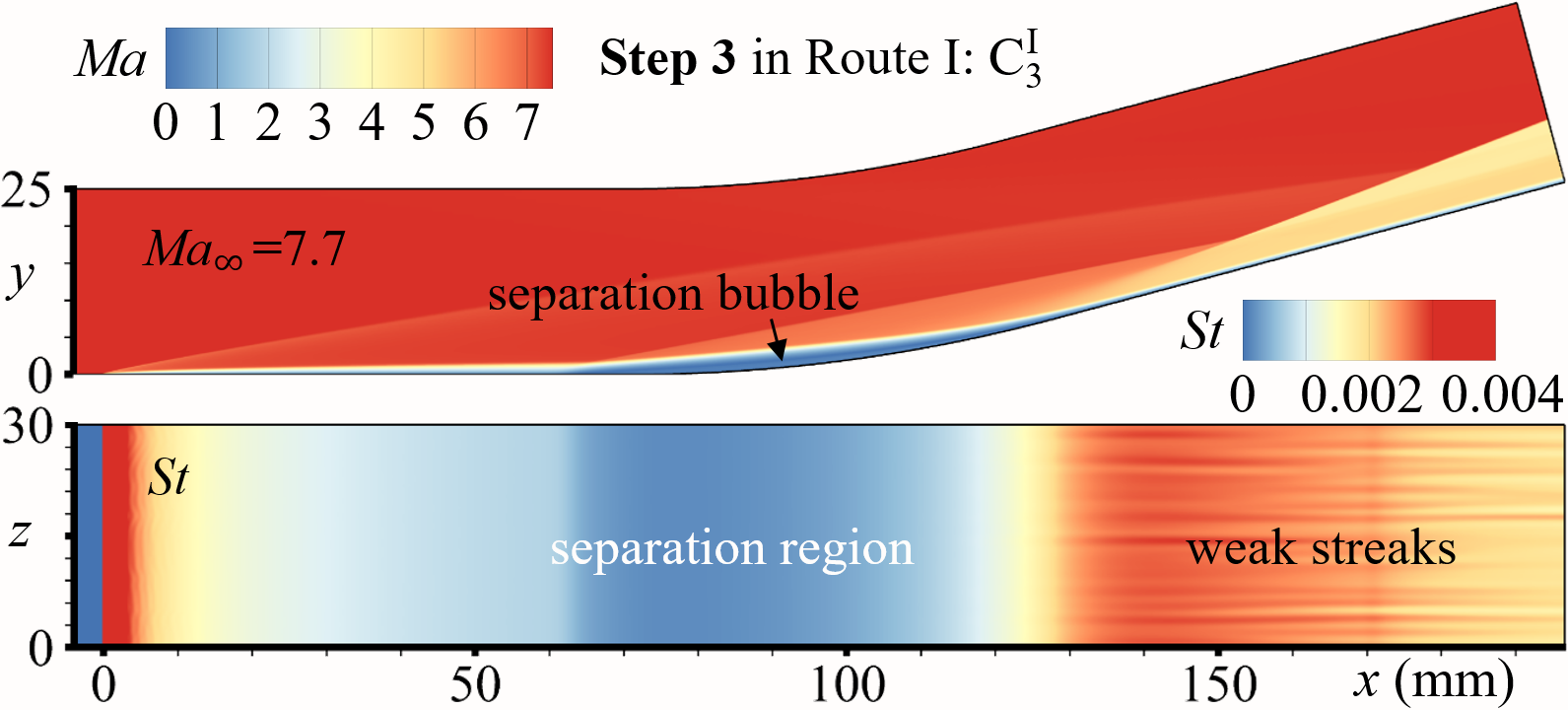}} 
	\subfigure[\label{subfig:C3_SA}]{\includegraphics[width = 0.48\columnwidth]{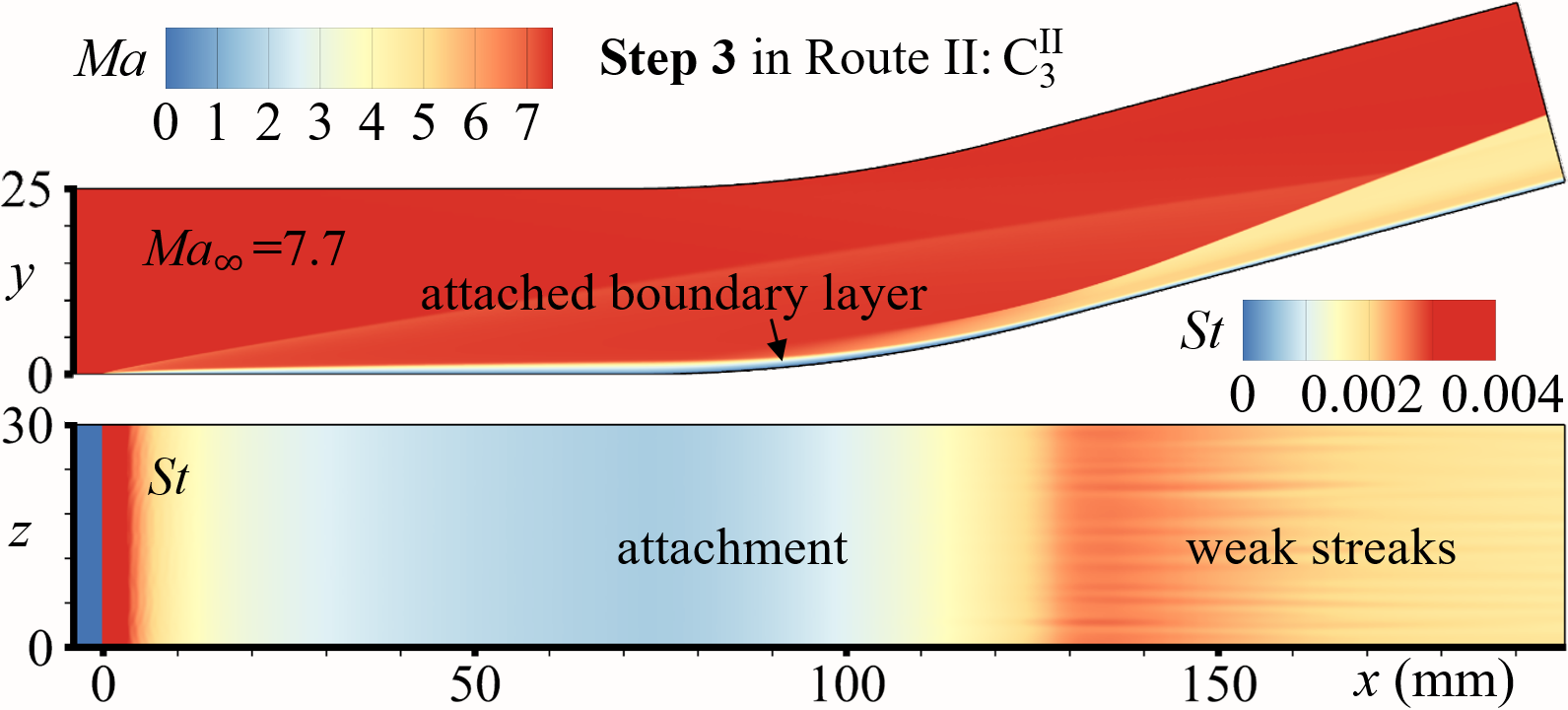}} 
	\caption{Terminuses of Route I and II. (a) C$_{1}^{\text{I}}$ and C$_{1}^{\text{II}}$ -separated; (b) C$_{4}^{\text{I}}$ and C$_{4}^{\text{II}}$-attached; (c) C$_{2}^{\text{I}}$-separated; (d) C$_{2}^{\text{II}}$-attached; (e) C$_{3}^{\text{I}}$-separated; and (f) C$_{3}^{\text{II}}$-attached.}\label{fig:DNS}
\end{figure}

 To accomplish Step 3 in Route II (adding viscosity), Euler equations are replaced with Navier-Stokes form, with the expectation of obtaining the stable attached states. The processes are denoted as C$_{\text{1}}^{\text{II,inv}}$ $\rightarrow$ C$_{\text{1}}^{\text{II}}$, C$_{\text{2}}^{\text{II,inv}}$ $\rightarrow$ C$_{\text{2}}^{\text{II}}$, C$_{\text{3}}^{\text{II,inv}}$ $\rightarrow$ C$_{\text{3}}^{\text{II}}$, and C$_{\text{4}}^{\text{II,inv}}$ $\rightarrow$ C$_{\text{4}}^{\text{II}}$, whose evolutionary histories of $x_{C_{f,min}}$ are respectively shown in Fig.\ref{subfig:CCR23_time}, \ref{subfig:CCR25_time}, \ref{subfig:CCR29_time}, and \ref{subfig:CCR30_time} (non-dimensional time $tu_{\infty}/L > 35$ for all cases, see the lower blue $x$-axes). $x_{C_{f,min}}$-variations in C$_{2}$, C$_{3}$, and C$_{4}$ are similar: being of only one branch, first moving downstream and then stabilizing after about $tu_{\infty}/L \approx 10$. This behavior implies  C$_{\text{2}}^{\text{II}}$, C$_{\text{3}}^{\text{II}}$, and C$_{\text{4}}^{\text{II}}$ all ultimately stabilize in attached states. However, in C$_{\text{1}}$ (the smallest curvature radius), the $x_{C_{f,min}}$ bifurcates into two branches at $tu_{\infty}/L \approx 3$. The lower and upper branches repectively converge to the $x_{S}$ and $x_{R}$ branches obtained from D$^{\text{I}}_{0}$ $\rightarrow$ C$_{1}^{\text{I}}$, implying C$_{\text{1}}^{\text{II}}$ finally stabilizes in the separated state. Flow fields of C$_{1}^{\text{II}}$, C$_{2}^{\text{II}}$, C$_{\text{3}}^{\text{II}}$, and C$_{\text{4}}^{\text{II}}$ are also shown in Fig. \ref{subfig:C1_SS}, \ref{subfig:C2_SA}, \ref{subfig:C3_SA}, and \ref{subfig:C4_SA}, respectively.  

\subsection {Contrast of the bistable states}
\label{subsec:Contrast}
\par The ultimate states of route I and II are listed in Table \ref{tab:Ultimate_state}. For configuration C$_{1}$, the ultimate states, C$_{1}^{\text{I}}$ and C$_{1}^{\text{II}}$, are both separated as shown in Fig. \ref{subfig:CCR23_time}; for configuration C$_{4}$, the ultimate states, C$_{4}^{\text{I}}$ and C$_{4}^{\text{II}}$, are both attached as shown in Fig. \ref{subfig:CCR29_time}. It needs to be emphasized that, for configurations C$_{2}$ and C$_{3}$ (Fig. \ref{subfig:CCR25_time} and \ref{subfig:CCR29_time}), both separated states (C$_{2}^{\text{I}}$ and C$_{3}^{\text{I}}$ in Fig. \ref{subfig:C2_SS} and \ref{subfig:C3_SS}) and attached (C$_{2}^{\text{II}}$ and C$_{3}^{\text{II}}$ in Fig. \ref{subfig:C2_SS} and \ref{subfig:C2_SA}) states can be stably established for the same boundary conditions, verifying the existence of the bistable states in CCR flows. The ultimate states of route I and II are listed in Table \ref{tab:Ultimate_state}. \textcolor{blue}{Also note that the weak streaks emerge in the downstream, implying the bistable states can resist disturbances of a certain intensity.}
\begin{table}
	\begin{center}
		\def~{\hphantom{0}}
		\begin{tabular}{lcccccccc}
			  & C$_{1}^{\text{I}}$ & C$_{1}^{\text{II}}$ & C$_{2}^{\text{I}}$ & C$_{2}^{\text{II}}$ & C$_{3}^{\text{I}}$ & C$_{3}^{\text{II}}$ & C$_{4}^{\text{I}}$ & C$_{4}^{\text{II}}$ \\[3pt]
			  & Separated & Separated & Separated & Attached & Separated & Attached & Attached & Attached \\
		\end{tabular}
		\caption{The ultimate states of the two routes in different configurations}
		\label{tab:Ultimate_state}
	\end{center}
\end{table}
\par
\begin{figure}
	\centering
	\subfigure[\label{subfig:double_region}]{\includegraphics[width = 0.48\columnwidth]{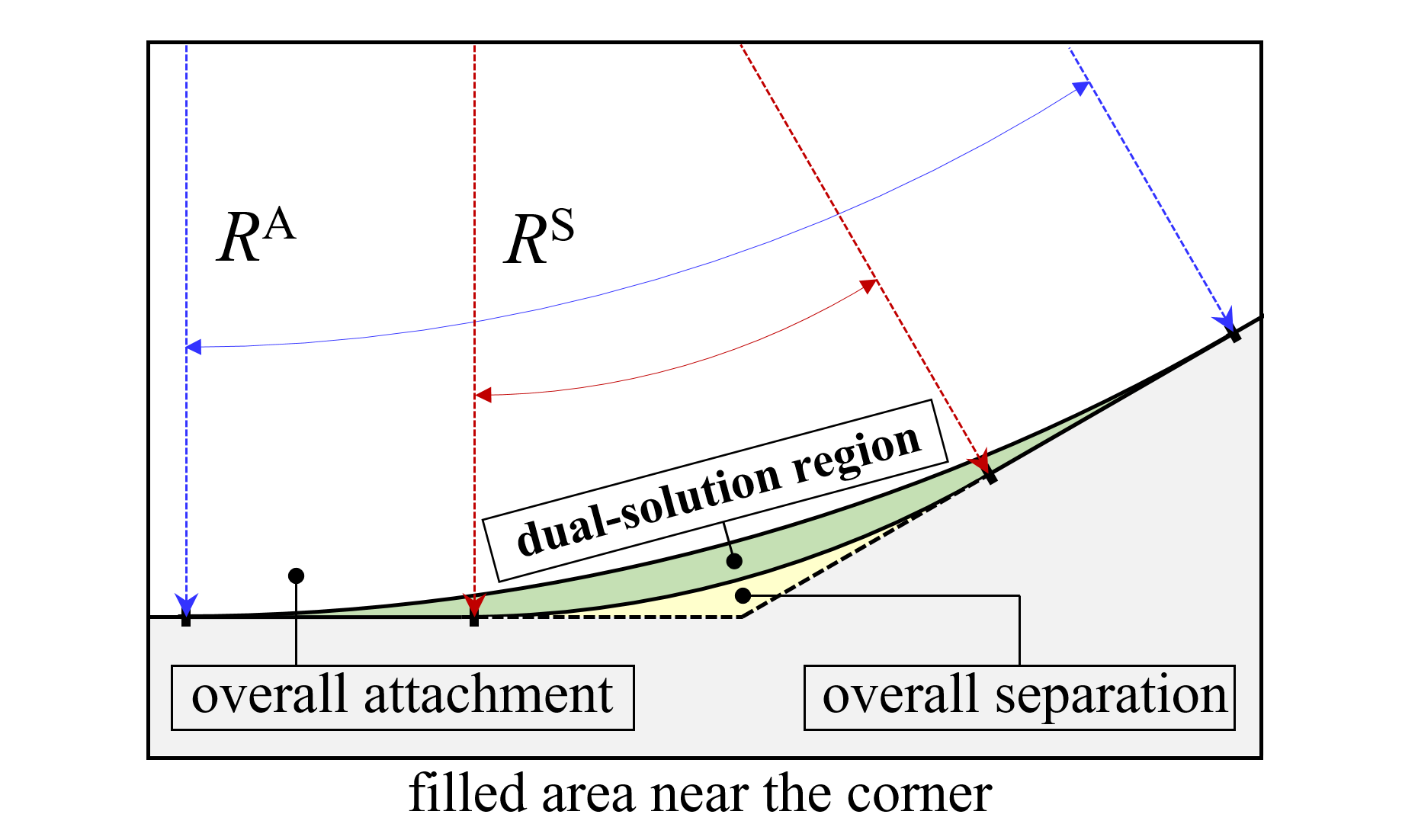}}
	\subfigure[\label{subfig:bistable_Cf}]{\includegraphics[width = 0.48\columnwidth]{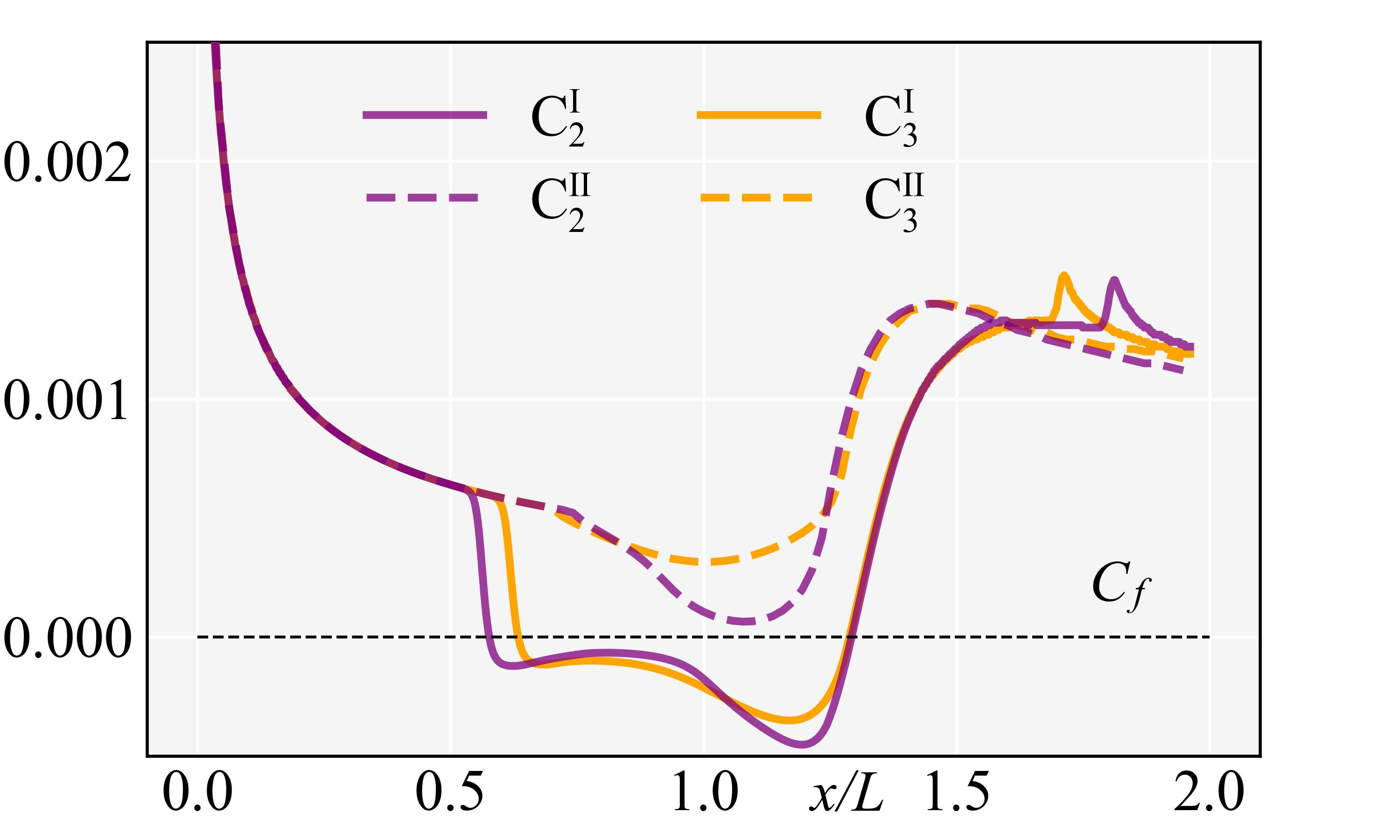}}
	\subfigure[\label{subfig:bistable_Cp}]{\includegraphics[width = 0.48\columnwidth]{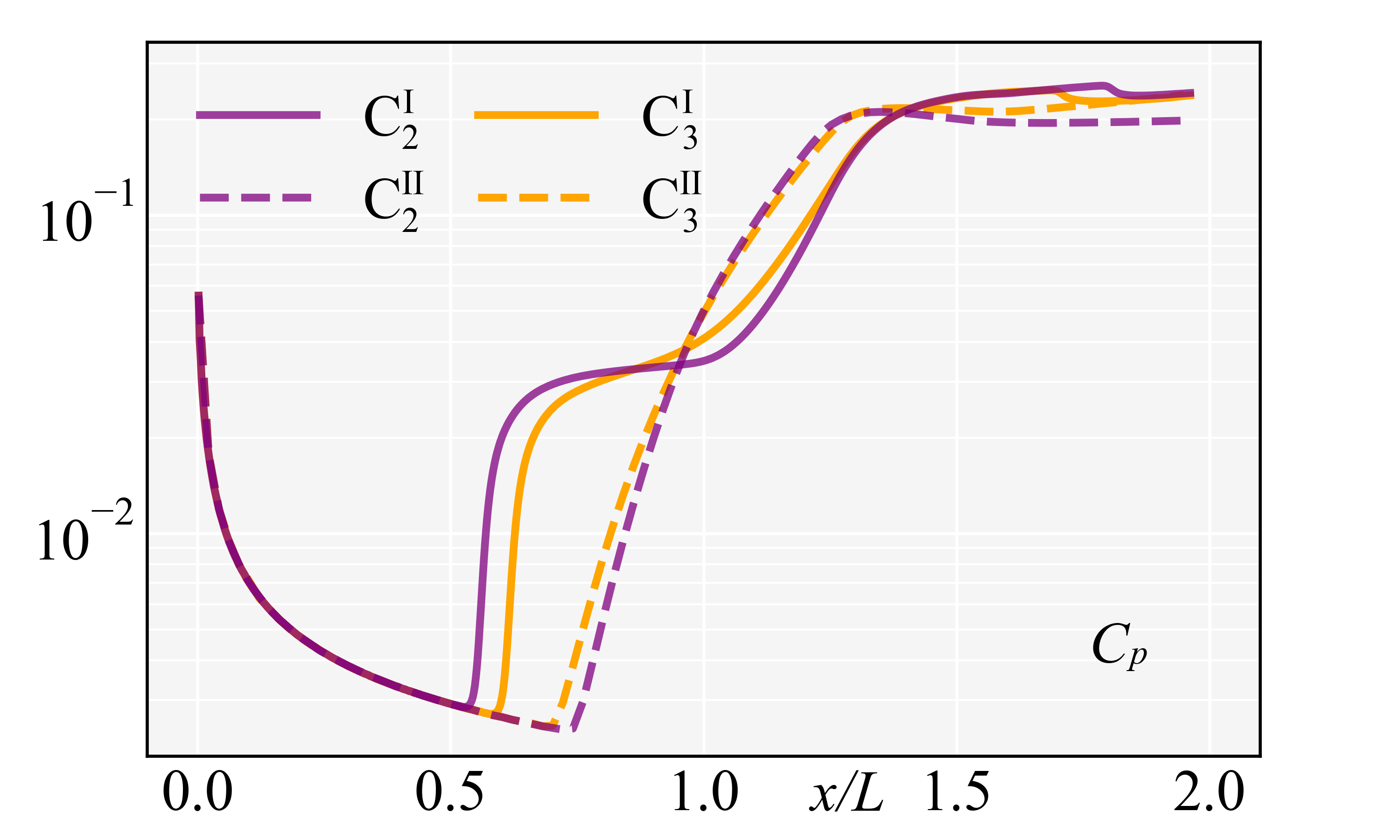}}
	\subfigure[\label{subfig:bistable_St}]{\includegraphics[width = 0.48\columnwidth]{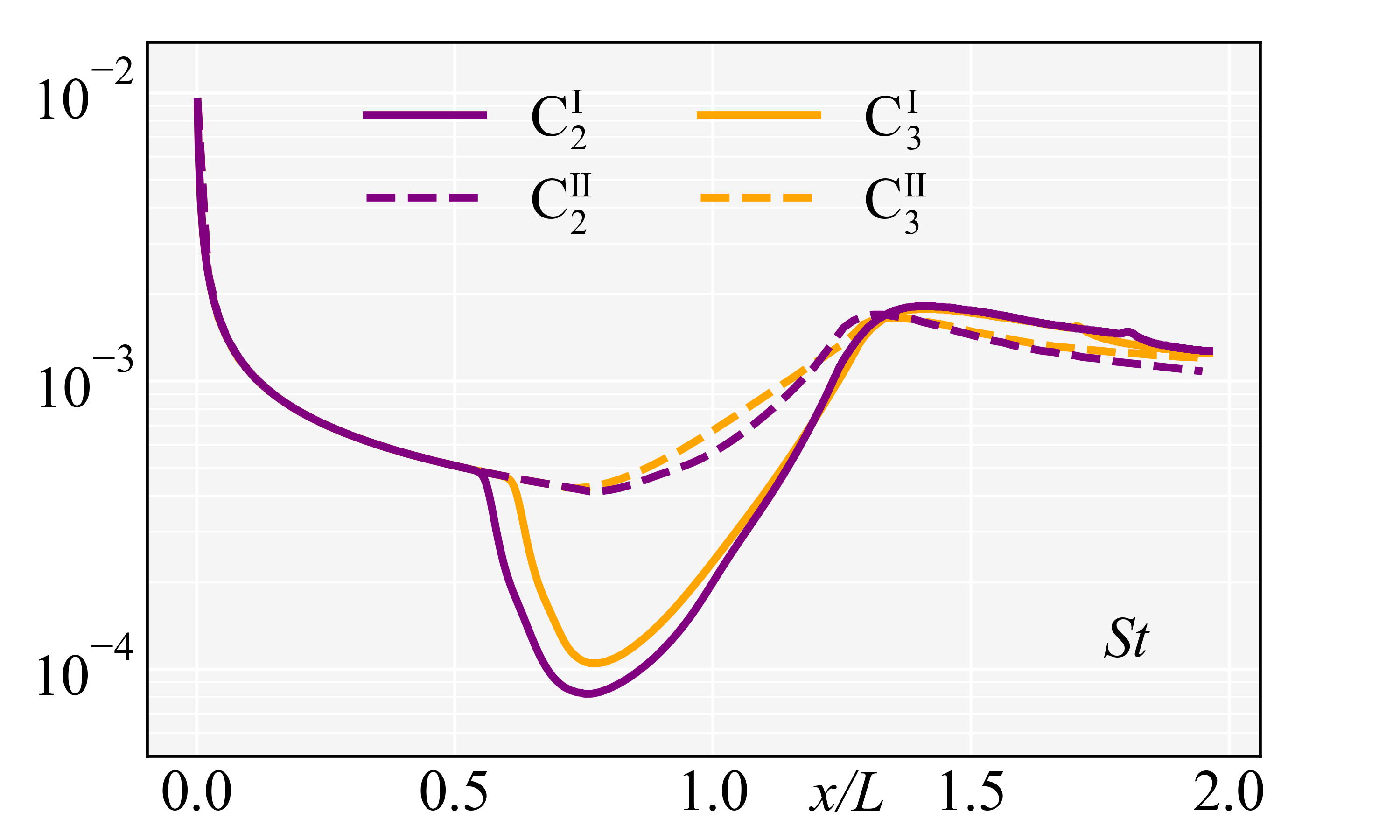}}
	\vspace{-2.5mm}
	\caption{(a) Schematic of the solution regions; distributions of (b) $C_f$; (c) $C_p$, and (d) $St$.}\label{fig:Bistable_Cp_Cf_St}
\end{figure}

 According to the ultimate states, the solution space of the flow field can be divided into three regions by two critical curvature radii, $R^{\text{A}}$ and $R^{\text{S}}$, as shown in Fig. \ref{subfig:double_region}: only stable attached states exist when $R > R^{\text{A}}$ (the overall attachment region); only stable separated states exist when $R < R^{\text{S}}$ (the overall separation region— the yellow area $\mathcal{A}_{\text{S}}$); both stable separate and attached states are physically possible when $R^{\text{A}} \leq R \leq R^{\text{S}}$ (the dual-solution region— the green area $\mathcal{A}_{\text{D}}$). The filled-area ratio $\eta$ is used to characterize the relative proportions of the dual-solution region, which is defined as
\begin{equation}
	\eta = \frac{\mathcal{A}_{\text{D}}}{\mathcal{A}_{\text{D}} + \mathcal{A}_{\text{S}}} = \frac{(R^{A})^2 [\text{tan}(\frac{\alpha}{2}) - \alpha] - (R^{S})^2 [ \text{tan}(\frac{\alpha}{2}) - \alpha]}{(R^{A})^2 [\text{tan}(\frac{\alpha}{2}) - \alpha]} = 1 - (\frac{R^{\text{S}}}{R^{\text{A}}})^2.
	\label{relative_proportions}
\end{equation}
In the present cases, $\eta \geq 25.69 \%$ with $R^{\text{S}} \leq R_{2} = 189.9$mm and $R^{\text{A}} \geq R_{3} = 220.3$mm. In other words, during the filling process, more than a quarter of the materials “break” the one-to-one corresponding relations from the configuration geometries to the flow states.
\par Furthermore, the aerothermodynamic characteristics including the distributions of spanwise-averaged $C_f$, $C_p$ and $St$ on C$_{2}$ and C$_{3}$ are shown in Fig. \ref{fig:Bistable_Cp_Cf_St}. For both separated states (C$_{2}^{\text{I}}$ and C$_{3}^{\text{I}}$) and attached (C$_{2}^{\text{II}}$ and C$_{3}^{\text{II}}$) states of C$_{2}$ and C$_{3}$, all the distributions of $C_f$ collapse in the upstream part of the flat plate, so do $C_p$ and $St$. However, in the downstream, the distributions are distinctly different. For the $C_f$ distributions, as shown in Fig. \ref{subfig:bistable_Cf}, C$_{2}^{\text{I}}$ and C$_{3}^{\text{I}}$ quickly drop to negatives after separation and sharply rise to positives after reattachment, and both of them have two minimal values in the seaparation regions; while both C$_{2}^{\text{II}}$ and C$_{3}^{\text{II}}$ stay positive, and have only one minimal value, severally. For $C_p$ distributions, as shown in Fig. \ref{subfig:bistable_Cp}, C$_{2}^{\text{I}}$ and C$_{3}^{\text{I}}$ present plateaus in the separation regions, rise sharply after reattachment and then reach peaks, severally; while both C$_{2}^{\text{II}}$ and C$_{3}^{\text{II}}$  continuously decrease till the starting points $M$ of the curved walls, and then rise in the form of isentropic compression. For the $St$ distributions, as shown in Fig. \ref{subfig:bistable_St}, C$_{2}^{\text{I}}$ and C$_{3}^{\text{I}}$ drop rapidly to near zero (\textit{O}($10^{-4}$)) after separation, and then increase rapidly to the peaks; while C$_{2}^{\text{II}}$ and C$_{3}^{\text{II}}$ slowly decrease till the starting points $M$, and then increase gradually. Additionally, in the downstream of the ramps, $C_f$, $C_p$, and $St$ of the separated states are all greater than those of the attached states.
\textcolor{blue}{\section {Robustness study}}
The phenomenon of bistability is somewhat counterintuitive, i.e., both separated and attached states can stably exist in C2 or C3 in the present cases, as shown in \ref{tab:Ultimate_state}. To confirm its existence in the in the real physical world with disturbances, we further investigate the robustness of this phenomenon. Specifically, the attached states, C$_{2}^{\text{II}}$ and C$_{3}^{\text{II}}$, are studied with some disturbances imparted to the upstream flow. 
\par The disturbances have the following form, a region of steady blowing and suction, referring to \cite{pirozzoli2004direct},
\begin{equation}
	v(x, z, t)=A u_{\infty} f(x) g(z) h(t), \quad x_a \leqslant x \leqslant x_b,
\end{equation}
with $A$ the amplitude of the disturbance, $u_{\infty}$ the freestream velocity, and
\begin{equation}
	\begin{aligned}
		& f(x)=4 \sin \theta(1-\cos \theta) / \sqrt{27}, \quad \theta=2 \pi\left(x-x_a\right) /\left(x_b-x_a\right), \\
		& g(z)=\sum_{l=1}^{l_{\max }} Z_l \sin \left[2 \pi l\left(z / L_z+\phi_l\right)\right], \quad \sum_{l=1}^{l_{\max }} Z_l=1, \quad Z_l=1.25 Z_{l+1}, \\
		& h(t) = 1.\\
		&
	\end{aligned}
\end{equation}
The locations $x_a = 15$mm and $x_b = 20$mm denote the beginning and the end of the disturbance region, respectively. Two amplitudes, $A = 1\%$ and $2\%$, are chosen to the investigate the influence of the disturbance strengths. $L_z = 30$mm is the spanwise width of the computational domain. $\phi_l$ is a random number ranging between 0 and 1. Three cases are considered, i.e., $A = 1\%$ \& $2\%$ for C$_{2}^{\text{II}}$ and $A = 2\%$ for C$_{3}^{\text{II}}$, all of which are simulated for $t u_{\infty} / L \geqslant 15$, and the results shown below are all at $t u_{\infty} / L = 15$.
\par
\begin{figure}
	\centering
	\subfigure[\label{subfig:A1_for_C2}]{\includegraphics[width = 0.7\columnwidth]{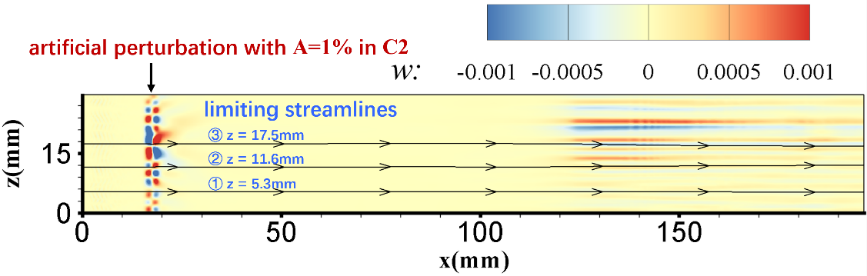}} 
	\subfigure[\label{subfig:A1_for_C2_Cf}]{\includegraphics[width = 0.5\columnwidth]{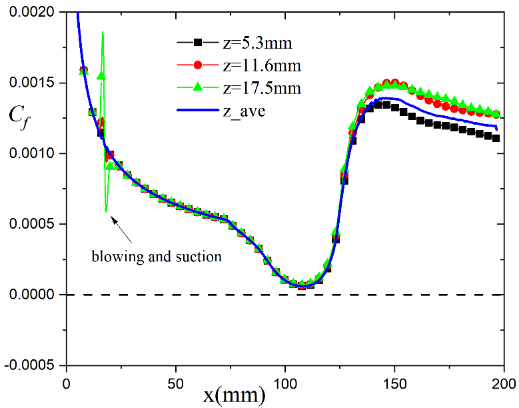}}
	\caption{Simulations of C$_{\text{1}}^{\text{II,inv}}$, C$_{\text{2}}^{\text{II,inv}}$, C$_{\text{3}}^{\text{II,inv}}$, and C$_{\text{4}}^{\text{II,inv}}$ to reproduce Step 2 in Route II. (a) flow fields colored by local Mach number in $x-y$ plane; (b) validation with Prandtl-Meyer relation of $C_p$.}\label{fig:A1_for_C2}
\end{figure}

\begin{figure}
	\centering
	\subfigure[\label{subfig:A2_for_C2}]{\includegraphics[width = 0.7\columnwidth]{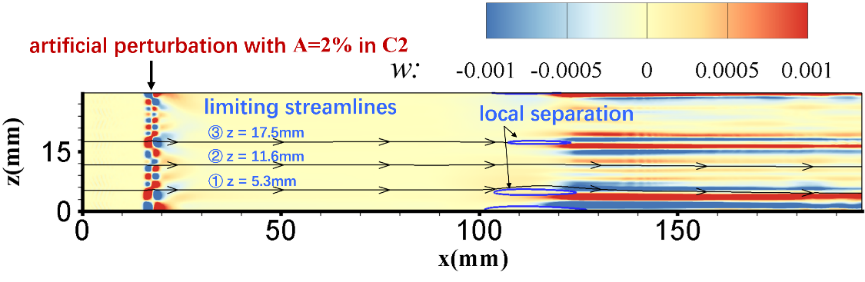}} 
	\subfigure[\label{subfig:A2_for_C2_Cf}]{\includegraphics[width = 0.5\columnwidth]{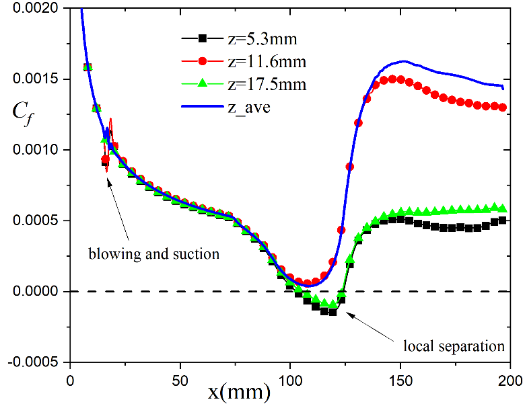}}
	\caption{Simulations of C$_{\text{1}}^{\text{II,inv}}$, C$_{\text{2}}^{\text{II,inv}}$, C$_{\text{3}}^{\text{II,inv}}$, and C$_{\text{4}}^{\text{II,inv}}$ to reproduce Step 2 in Route II. (a) flow fields colored by local Mach number in $x-y$ plane; (b) validation with Prandtl-Meyer relation of $C_p$.}\label{fig:A2_for_C2}
\end{figure}

 For C$_{2}^{\text{II}}$ (the smaller curvature radius R$\approx 189.9$mm), the weaker disturbance ($A = 1\%$) can not perturb the attached state into a separated state, as shown in Fig. \ref{fig:A1_for_C2}, while the stronger disturbance ($A = 2\%$) make the downstream boundary locally separate at $100$mm $\leqslant x \leqslant$ $125$mm, as shown in Fig. \ref{fig:A2_for_C2}. This implies that the stable attached state in C2 can at least resist a disturbance with the amplitude of $1\% u_{\infty}$, and for a fixed shape, the weaker the disturbance, the less likely the attached state is to be broken.
\par
\begin{figure}
	\centering
	\subfigure[\label{subfig:A2_for_C3}]{\includegraphics[width = 0.7\columnwidth]{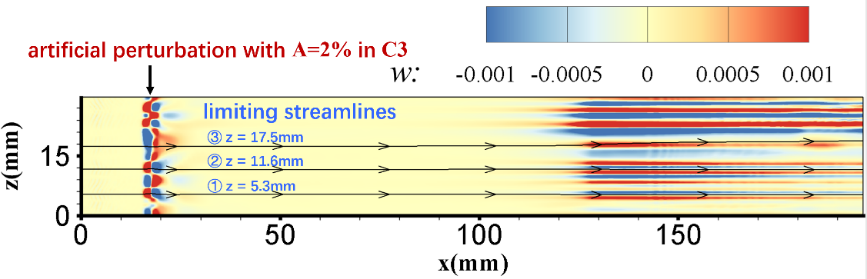}} 
	\subfigure[\label{subfig:A2_for_C3_Cf}]{\includegraphics[width = 0.5\columnwidth]{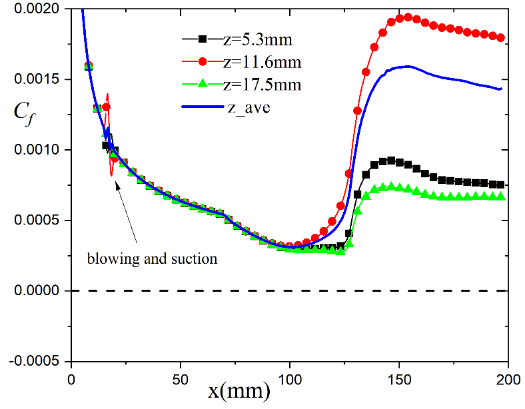}}
	\caption{Simulations of C$_{\text{1}}^{\text{II,inv}}$, C$_{\text{2}}^{\text{II,inv}}$, C$_{\text{3}}^{\text{II,inv}}$, and C$_{\text{4}}^{\text{II,inv}}$ to reproduce Step 2 in Route II. (a) flow fields colored by local Mach number in $x-y$ plane; (b) validation with Prandtl-Meyer relation of $C_p$.}\label{fig:A2_for_C3}
\end{figure}

 For C$_{3}^{\text{II}}$ (the larger curvature radius R$\approx 220.3$mm), even the stronger disturbance ($A = 2\%$) can not perturb the attached state into a separated state, as shown in Fig. \ref{fig:A2_for_C3}. In fact, it can be seen that the $C_f$ on the curved wall in C$_{3}$ is larger (a greater distance from $C_f = 0$) than that of C$_{2}$, which can resist a stronger disturbance. This implies that, for a certain disturbance, the larger the curvature radius, the more robust the attached state.
\par The above results imply that the bistable states can resist disturbances of a certain intensity; the larger the curvature radius, the more robust the attached state; the smaller the curvature radius, the more robust the separated state.
\section{Conclusions}
\label{sec:Conclusions}
\par The CCR flows' bistability is conjectured through a thought experiment and verified using the 3D DNSs. The dual-solution region in the present cases accounts for more than 25 percent of the total filling area. The differences of bistable aerothermodynamic are compared. As an intrinsic property, CCR's bistability originates from the bifurcation characteristics of Navier-Stokes equations, and its specific presentation process can be diverse, corresponding to the hystereses induced by different parameter variations, such as $Ma_{\infty}$, $\alpha$, and $T_{w}$- variations, and some 2D cases have been reported \citep{hu2020bistable,zhou2021mechanism}.
\par Next, we will design wind tunnel experiments to show the bistable states of CCR flows, which is more interesting and challenging. After all, it took more than 100 years from Mach's discovery \citep{mach1878uber} of two SW reflection patterns to the observation of their bistable states in the wind tunnel \citep{chpoun1995reconsideration}. In some sense, there is systematic similarity between CCR flows and SW reflections: local subsonic regions (separation bubbles in CCR flows, and subsonic regions behind Mach stems in SW reflections) exist in the global supersonic/hypersonic flows, making the flow systems have both hyperbolic and elliptic characteristics. Therefore, the geometrical parameters, curvature radius $R$ and wedge angle $\theta_{w}$, play the similar role in the bistabilities of CCR flows and SW reflections, respectively.
\par SBLIs, represented by CCR flows, and Shock-Shock interactions (SSIs), represented by SW reflections \citep{ben2001hysteresis}, often dominate the complex flow in supersonic/hypersonic flight together. Therefore, more complex multistable shock patterns will be formed when multistable states of SBLIs and SSIs interact with each other, and the resultant complex aerothermodynamic characteristics need to be paid more attention in the future. 
\section*{Acknowledgment}
This work is supported by the National Key R \& D Program of China (Grant No.2019YFA0405300). We look forward to receiving helpful comments from reviewers.
\vspace{2.5mm}
\par \hspace{-4.5mm} \textbf{Declaration of interests.} The authors report no conflict of interest.
\vspace{-2.5mm}
\bibliographystyle{jfm}
\bibliography{jfm2}

\begin{thebibliography}{30}
\expandafter\ifx\csname natexlab\endcsname\relax\def\natexlab#1{#1}\fi
\def\au#1{#1} \def\ed#1{#1} \def\yr#1{#1}\def\at#1{#1}\def\jt#1{\textit{#1}}
  \def\bt#1{#1}\def\bvol#1{\textbf{#1}} \def\vol#1{#1} \def\pg#1{#1}
  \def\publ#1{#1}\def\arxiv#1{#1}\def\org#1{#1}\def\st#1{\textit{#1}}

\bibitem[{Babinsky} \& {Harvey}(2011)]{babinsky2011shock}
{\sc \au{{Babinsky}, H.} \& \au{{Harvey}, J.~K.}} \yr{2011} {\em Shock
  Wave-Boundary-Layer Interactions\/}.

\bibitem[Ben-Dor {\em et~al.\/}(2001)Ben-Dor, Elperin \&
  Chpoun]{ben2001hysteresis}
{\sc \au{Ben-Dor, G., Vasiliev~EI.}, \au{Elperin, T.} \& \au{Chpoun, A.}}
  \yr{2001}  \at{Hysteresis phenomena in the interaction process of conical
  shock waves: experimental and numerical investigations}.  \jt{J. Fluid Mech.}
   \bvol{448}.

\bibitem[{Cao} {\em et~al.\/}(2021){Cao}, {Hao}, {Klioutchnikov}, {Olivier} \&
  {Wen}]{cao2021unsteady}
{\sc \au{{Cao}, S.~B.}, \au{{Hao}, J.}, \au{{Klioutchnikov}, I.},
  \au{{Olivier}, H.} \& \au{{Wen}, C.~Y.}} \yr{2021}  \at{Unsteady effects in a
  hypersonic compression ramp flow with laminar separation}.  \jt{J. Fluid
  Mech.}  \bvol{912}.

\bibitem[{Chapman} {\em et~al.\/}(1958){Chapman}, {Kuehn} \&
  {Larson}]{chapman1958investigation}
{\sc \au{{Chapman}, D.~R.}, \au{{Kuehn}, D.~M.} \& \au{{Larson}, H.~K.}}
  \yr{1958}  \at{Investigation of separated flows in supersonic and subsonic
  streams with emphasis on the effect of transition} .

\bibitem[Chpoun \& Ben-Dor(1995)]{chpoun1995reconsideration}
{\sc \au{Chpoun, A., Passerel D. Li~H.} \& \au{Ben-Dor, G.}} \yr{1995}
  \at{Reconsideration of oblique shock wave reflections in steady flows. part
  1. experimental investigation}.  \jt{J. Fluid Mech.}  \bvol{301},
  \pg{19--35}.

\bibitem[{Chuvakhov} {\em et~al.\/}(2017){Chuvakhov}, {Borovoy}, {Egorov},
  {Radchenko}, {Olivier} \& {Roghelia}]{chuvakhov2017effect}
{\sc \au{{Chuvakhov}, P.~V.}, \au{{Borovoy}, V.~Ya.}, \au{{Egorov}, I.~V.},
  \au{{Radchenko}, V.~N.}, \au{{Olivier}, H.} \& \au{{Roghelia}, A.}} \yr{2017}
   \at{Effect of small bluntness on formation of görtler vortices in a
  supersonic compression corner flow}.  \jt{J. Appl. Mech. Tech. Phys.}
  \bvol{58}~(6),  \pg{975--989}.

\bibitem[{Edney}(1968)]{edney1968anomalous}
{\sc \au{{Edney}, B.}} \yr{1968}  \at{Anomalous heat transfer and pressure
  distributions on blunt bodies at hypersonic speeds in the presence of an
  impinging shock} .

\bibitem[{Fu} {\em et~al.\/}(2021){Fu}, {Karp}, {Bose}, {Moin} \&
  {Urzay}]{fu2021shock}
{\sc \au{{Fu}, L.}, \au{{Karp}, M.}, \au{{Bose}, S.~T.}, \au{{Moin}, P.} \&
  \au{{Urzay}, J.}} \yr{2021}  \at{Shock-induced heating and transition to
  turbulence in a hypersonic boundary layer}.  \jt{J. Fluid Mech.}  \bvol{909}.

\bibitem[{Gai} \& {Khraibut}(2019)]{gai2019hypersonic}
{\sc \au{{Gai}, S.~L.} \& \au{{Khraibut}, A.}} \yr{2019}  \at{Hypersonic
  compression corner flow with large separated regions}.  \jt{J. Fluid Mech.}
  \bvol{877},  \pg{471--494}.

\bibitem[{Ganapathisubramani} {\em et~al.\/}(2009){Ganapathisubramani},
  {Clemens} \& {Dolling}]{ganapathisubramani2009low}
{\sc \au{{Ganapathisubramani}, B.}, \au{{Clemens}, N.~T.} \& \au{{Dolling},
  D.~S.}} \yr{2009}  \at{Low-frequency dynamics of shock-induced separation in
  a compression ramp interaction}.  \jt{J. Fluid Mech.}  \bvol{636},
  \pg{397--425}.

\bibitem[{Gumand}(1959)]{gumand1959on}
{\sc \au{{Gumand}, W.~J.}} \yr{1959}  \at{On the plateau and peak pressure of
  regions of pure laminar and fully turbulent separation in two-dimensional
  supersonic flow}.  \jt{J. Aerosp. Sci.}  \bvol{26}~(1),  \pg{56--56}.

\bibitem[{Helm} {\em et~al.\/}(2021){Helm}, {Martín} \&
  {Williams}]{helm2021characterization}
{\sc \au{{Helm}, C.~M.}, \au{{Martín}, M.~P.} \& \au{{Williams}, O.~J.H.}}
  \yr{2021}  \at{Characterization of the shear layer in separated
  shock/turbulent boundary layer interactions}.  \jt{J. Fluid Mech.}
  \bvol{912}.

\bibitem[{Hu} {\em et~al.\/}(2017){Hu}, {Bi}, {Li} \& {She}]{hu2017beta}
{\sc \au{{Hu}, Y.~C.}, \au{{Bi}, W.~T.}, \au{{Li}, S.~Y.} \& \au{{She}, Z.~S.}}
  \yr{2017}  \at{beta-distribution for reynolds stress and turbulent heat flux
  in relaxation turbulent boundary layer of compression ramp}.  \jt{Sci. China:
  Phys., Mech. Astron.}  \bvol{60}~(12),  \pg{124711}.

\bibitem[{Hu} {\em et~al.\/}(2020){Hu}, {Zhou}, {Wang}, {Yang} \&
  {Tang}]{hu2020bistable}
{\sc \au{{Hu}, Y.~C.}, \au{{Zhou}, W.~F.}, \au{{Wang}, G.}, \au{{Yang}, Y.~G.}
  \& \au{{Tang}, Z.~G.}} \yr{2020}  \at{Bistable states and separation
  hysteresis in curved compression ramp flows}.  \jt{Phys. Fluids}
  \bvol{32}~(11),  \pg{113601}.

\bibitem[{Hung}(1973)]{hung1973interference}
{\sc \au{{Hung}, F.~T.}} \yr{1973} Interference heating due to shock wave
  impingement on laminar boundary layers.  \bt{In {\em 6th Fluid and
  PlasmaDynamics Conference\/}}.

\bibitem[Jiang \& Shu(1996)]{jiang1996efficient}
{\sc \au{Jiang, Guang-Shan} \& \au{Shu, Chi-Wang}} \yr{1996}  \at{Efficient
  implementation of weighted eno schemes}.  \jt{Journal of computational
  physics}  \bvol{126}~(1),  \pg{202--228}.

\bibitem[{Mach}(1878)]{mach1878uber}
{\sc \au{{Mach}, E.}} \yr{1878}  \at{Uber den verlauf von funkenwellen in der
  ebene und im raume}.  \jt{Sitzungsbr. Akad. Wiss. Wien}  \bvol{78},
  \pg{819--838}.

\bibitem[{Neiland}(1969)]{neiland1969theory}
{\sc \au{{Neiland}, V.~Y.}} \yr{1969}  \at{Theory of laminar boundary layer
  separation in supersonic flow}.  \jt{Fluid Dyn.}  \bvol{4}~(4).

\bibitem[Pirozzoli {\em et~al.\/}(2004)Pirozzoli, Grasso \&
  Gatski]{pirozzoli2004direct}
{\sc \au{Pirozzoli, Sergio}, \au{Grasso, F} \& \au{Gatski, TB}} \yr{2004}
  \at{Direct numerical simulation and analysis of a spatially evolving
  supersonic turbulent boundary layer at m= 2.25}.  \jt{Physics of fluids}
  \bvol{16}~(3),  \pg{530--545}.

\bibitem[{Roghelia} {\em et~al.\/}(2017{\natexlab{{\em a\/}}}){Roghelia},
  {Chuvakhov}, {Olivier} \& {Egorov}]{roghelia2017experimental}
{\sc \au{{Roghelia}, A.}, \au{{Chuvakhov}, P.~V.}, \au{{Olivier}, H.} \&
  \au{{Egorov}, I.}} \yr{2017{\natexlab{{\em a\/}}}}  \at{Experimental
  investigation of görtler vortices in hypersonic ramp flows behind sharp and
  blunt leading edges}.  \jt{AIAA paper}  \pg{p. 3463}.

\bibitem[{Roghelia} {\em et~al.\/}(2017{\natexlab{{\em b\/}}}){Roghelia},
  {Olivier}, {Egorov} \& {Chuvakhov}]{Roghelia2017ExperimentalIO}
{\sc \au{{Roghelia}, A.}, \au{{Olivier}, H.}, \au{{Egorov}, I.~V.} \&
  \au{{Chuvakhov}, P.~V.}} \yr{2017{\natexlab{{\em b\/}}}}  \at{Experimental
  investigation of g{\"o}rtler vortices in hypersonic ramp flows}.  \jt{Exp.
  Fluids}  \bvol{58},  \pg{1--15}.

\bibitem[{Simeonides} \& {Haase}(1995)]{simeonides1995experimental}
{\sc \au{{Simeonides}, G.} \& \au{{Haase}, W.}} \yr{1995}  \at{Experimental and
  computational investigations of hypersonic flow about compression ramps}.
  \jt{J. Fluid Mech.}  \bvol{283}~(-1),  \pg{17--42}.

\bibitem[{Stewartson} \& {Williams}(1969)]{stewartson1969self}
{\sc \au{{Stewartson}, K.} \& \au{{Williams}, P.~G.}} \yr{1969}
  \at{Self-induced separation}.  \jt{Proc. R. Soc. London, Ser. A}
  \bvol{312}~(1509).

\bibitem[{Tang} {\em et~al.\/}(2021){Tang}, {Wang}, {Hu}, {Zhou}, {Xie} \&
  {Yang}]{tang2021aerothermodynamic}
{\sc \au{{Tang}, M.~Z.}, \au{{Wang}, G.}, \au{{Hu}, Y.~C.}, \au{{Zhou}, W.~F.},
  \au{{Xie}, Z.~X.} \& \au{{Yang}, Y.~G.}} \yr{2021}  \at{Aerothermodynamic
  characteristics of hypersonic curved compression ramp flows with bistable
  states}.  \jt{Phys. Fluids}  \bvol{33}~(12),  \pg{126106}.

\bibitem[{Tao} {\em et~al.\/}(2014){Tao}, {Fan} \& {Zhao}]{tao2014viscous}
{\sc \au{{Tao}, Y.}, \au{{Fan}, X.~Q.} \& \au{{Zhao}, Y.~L.}} \yr{2014}
  \at{Viscous effects of shock reflection hysteresis in steady supersonic
  flows}.  \jt{J. Fluid Mech.}  \bvol{759},  \pg{134--148}.

\bibitem[{Tong} {\em et~al.\/}(2017){Tong}, {Li}, {Duan} \&
  {Yu}]{tong2017direct}
{\sc \au{{Tong}, F.~L.}, \au{{Li}, X.~L.}, \au{{Duan}, Y.H.} \& \au{{Yu},
  C.~P.}} \yr{2017}  \at{Direct numerical simulation of supersonic turbulent
  boundary layer subjected to a curved compression ramp}.  \jt{Phys. Fluids}
  \bvol{29}~(12),  \pg{125101}.

\bibitem[{Wang} {\em et~al.\/}(2019){Wang}, {Wang}, {Sun}, R.{Yang}, {Zhao} \&
  {Hu}]{wang2019amplification}
{\sc \au{{Wang}, Q.~C.}, \au{{Wang}, Z.~G.}, \au{{Sun}, M.~B.}, \au{R.{Yang}},
  \au{{Zhao}, Y.X.} \& \au{{Hu}, Z.~W.}} \yr{2019}  \at{The amplification of
  large-scale motion in a supersonic concave turbulent boundary layer and its
  impact on the mean and statistical properties}.  \jt{J. Fluid Mech.}
  \bvol{863},  \pg{454--493}.

\bibitem[{White} \& {Majdalani}(2006)]{white2006viscous}
{\sc \au{{White}, F.M.} \& \au{{Majdalani}, J.}} \yr{2006} {\em Viscous fluid
  flow\/}, ,  \vol{vol.~3}.  \publ{McGraw-Hill New York}.

\bibitem[{Xu} {\em et~al.\/}(2021){Xu}, {Wang}, {Wan}, {Yu}, {Li} \&
  {Chen}]{xu2021effect}
{\sc \au{{Xu}, D.~H.}, \au{{Wang}, J.~C.}, \au{{Wan}, M.~P.}, \au{{Yu}, C.~P.},
  \au{{Li}, X.~L.} \& \au{{Chen}, S.~Y.}} \yr{2021}  \at{Effect of wall
  temperature on the kinetic energy transfer in a hypersonic turbulent boundary
  layer}.  \jt{J. Fluid Mech.}  \bvol{929}.

\bibitem[{Zhou} {\em et~al.\/}(2021){Zhou}, {Hu}, {Tang}, {Wang}, {Fang} \&
  {Yang}]{zhou2021mechanism}
{\sc \au{{Zhou}, W.~F.}, \au{{Hu}, Y.~C.}, \au{{Tang}, M.~Z.}, \au{{Wang}, G.},
  \au{{Fang}, M.} \& \au{{Yang}, Y.~G.}} \yr{2021}  \at{Mechanism of separation
  hysteresis in curved compression ramp}.  \jt{Phys. Fluids}  \bvol{33}~(10),
  \pg{106108}.

\end{thebibliography}


\end{document}